\documentclass[aps,pra,twocolumn,amsmath,amssymb,floatfix,longbibliography,10pt]{revtex4-2}

\usepackage[english]{babel}
\usepackage[utf8]{inputenc}
\usepackage{graphicx}
\usepackage{xparse}
\usepackage{etoolbox}
\usepackage{mathtools}
\usepackage{subdepth}
\usepackage{physics}
\usepackage{hyperref}
\usepackage{glossaries}
\usepackage[capitalise,english]{cleveref}
\usepackage{csquotes}
\usepackage{dsfont}
\usepackage{xparse}
\usepackage{tabularx}
\usepackage{xcolor}

\usepackage{float}
\makeatletter
  \usepackage{xspace}
  \usepackage{xpunctuate}

  \appto{\appendix}{%
    \@ifstar{\def\theequation@prefix{A.}}%
            {}}
\makeatother

\graphicspath{{Figures/}}

\crefname{equation}{Eq.}{Eqs.}
\Crefname{equation}{Eq.}{Eqs.}

\crefname{section}{Sec.}{Sec.s}
\Crefname{section}{Section}{Sections}

\newcommand{\smalldagger}{\text{\raisebox{-2pt}{$\dagger$}}}
\newcommand{\smallsquare}{\text{\raisebox{-2pt}{$2$}}}

\begin{document}

\begin{abstract}	
	
We derive a Markovian master equation for a linearly driven dissipative quantum harmonic oscillator, valid for generic driving beyond the adiabatic limit. We solve this quantum master equation for arbitrary Gaussian initial states and investigate its departure from the adiabatic master equation in the regime of fast driving. We concretely examine the behavior of dynamical variables, such as position and momentum, as well as of thermodynamic quantities, such as energy and entropy. We additionally  study the influence of the nonequilibrium driving on the quantum coherence of the oscillator in the instantaneous energy eigenbasis. We further analyze the approach to the adiabatic limit and the relaxation to the instantaneous {steady} state as a function of the driving speed.

\end{abstract}

\title{Nonadiabatic master equation for a linearly driven harmonic oscillator}
\author{Sinan Altinisik}
\affiliation{Institute for Theoretical Physics I, University of Stuttgart, D-70550 Stuttgart, Germany}
\author{Eric Lutz}
\affiliation{Institute for Theoretical Physics I, University of Stuttgart, D-70550 Stuttgart, Germany}

\maketitle

\section{Introduction}
\label{sec:introduction}

Markovian master equations are an  essential tool in the study of open quantum systems. They provide a powerful, yet approximate, description of the time evolution of the reduced density operator of  systems weakly coupled to their  environments \cite{bre02,gar04,ali07,riv12}. They allow one to investigate  the dynamics of both diagonal density matrix elements (populations), that are involved in relaxation and thermalization processes, and of nondiagonal density matrix elements (coherences), that are connected to dephasing and decoherence phenomena. In view of their versatility, they have been successfully employed  in many  different  areas  in the past decades, ranging from quantum
optics \cite{car93}, condensed-matter physics \cite{wei08} and nonequilibrium statistical mechanics \cite{zwa01}, to quantum information theory \cite{nie00} and, more recently, quantum thermodynamics \cite{kur21}.

In many applications, the  open quantum system of interest is  subjected to an additional time-dependent external driving. A case in point is an atom coupled to a classical electric field  or a driven field mode  in a microwave cavity \cite{bre02,gar04}. In this  situation, the  derivation of the Markovian master equation is more involved than for static problems \cite{dav78}. For weak driving in the weak coupling limit,  the Hamiltonian is often simply  replaced by the time-dependent Hamiltonian in the otherwise undriven  master equation \cite{bre02,gar04} (see also Refs.~\cite{ali79,ali06,gev95,koh99}). This approach has been shown to lead to inconsistencies, in particular  with the second law of thermodynamics \cite{gev95,koh99,szc13}. The instance of periodically driven open quantum  systems (fast and slow) can be conveniently analyzed using Floquet theory \cite{zer94,bre97,kam11,ali12,dit98}. On the other hand, a consistent framework for adiabatically evolving open quantum  systems (not necessarily periodically driven),   has been formulated recently \cite{alb12,yip18}, and applied, for example, to quantum annealing \cite{boi13,pud14}. However, despite its central importance, especially for the analysis of far-from-equilibrium quantum phenomena, the case of aperiodic arbitrary  fast driving, beyond the adiabatic regime, has only received little attention so far \cite{bul15,yam17,dan18,moz20,meg24}.

We here consider a paradigmatic model consisting of a linearly driven quantum harmonic oscillator weakly coupled to a bosonic bath \cite{gar04}. We derive a Markovian master equation for its reduced density operator valid for arbitrary driving speed, and provide an explicit solution for generic initial Gaussian states. We investigate in detail the dynamics of position and momentum,  mean energy and von Neumann entropy, as well as a measure of quantum coherence in the instantaneous energy eigenbasis of the driven oscillator for the concrete example of a linear ramp. We further examine the transition to the adiabatic limit for slow external driving \cite{alb12,yip18} and contrast their different features. We, moreover, analyze the relaxation to the instantaneous steady state which, in general, is not given by a thermal Gibbs state.

The paper is organized as follows. \Cref{sec:derivation} begins with a derivation of the nonadiabatic master equation for the driven quantum harmonic oscillator in the Born-Markov approximation. \Cref{sec:ad_and_wd_me} discusses the transition to the corresponding  adiabatic master equation in the limit of slow driving. The case of a linear driving protocol is then treated in detail in \cref{sec:ramp}. The solution of the nonadiabatic master equation for generic initial Gaussian states by means of the adjoint master equation is further presented in \cref{sec:adjoint_me_and_obs}, and applied to position and momentum in \cref{sec:pm} and to energy and entropy  in \cref{sec:E_and_S}.  The properties of the quantum coherence in the instantaneous energy eigenbasis is additionally examined  in \cref{sec:coherence}, whereas the approach to the instantaneous steady state, and its deviation from the instantaneous Gibbs state, are analyzed in \cref{sec:iss}.

\section{Derivation of the Master equation}
\label{sec:derivation}

We consider a linearly driven, damped quantum harmonic oscillator with time-dependent Hamiltonian \cite{gar04}
\begin{equation}
	H_\text{S}(t)=\hbar\omega\left(n+\frac{1}{2}\right)-{\hbar\omega\frac{\lambda(t)}{2}}(a+a^\dagger)+\hbar\omega\frac{\lambda^2(t)}{4},
	\label{eq:H_S}
\end{equation}
where $\omega$ is the frequency and $\lambda(t)$ is a time-dependent driving parameter. The variable $\lambda(t)$ physically describes the shift of the equilibrium point of the potential in units of the zero-point fluctuation, $x_0=\sqrt{\hbar/2m\omega}$ \cite{HSrxp}. {For convenience, we set $\lambda(0)=0$.} As usual, $a$ ($a^{\dagger}$) denotes the annihilation (creation) operator and $n= a^\dagger a$ is the number operator. The last term in Eq.~\eqref{eq:H_S} ensures that the minimum of the  potential remains equal to zero; it has no influence on the dynamics. The oscillator is weakly coupled, via the interaction Hamiltonian $H_\text{I}=a^\dagger\otimes b+a\otimes b^\dagger$, to a bosonic bath with Hamiltonian $H_\text{B}=\sum_k\hbar\omega_kb_k^\dagger b_k$, where $b_k$ ($b_k^{\dagger}$) is the annihilation (creation) operator of the $k$-th bath mode with frequency $\omega_k$, and $b=\sum_kg_kb_k$ with  coupling constants $g_k$.
We assume that the bath is initially in a thermal state, ${\rho_\text{B}^\beta}=\exp(-\beta H_\text{B})/{Z_\text{B}}$, at inverse temperature $\beta$ with  partition function {$Z_\text{B}=\Tr_\text{B}[\exp(-\beta H_\text{B})]$}. 
 
The starting point of our analysis is the standard Born-Markov expression for the reduced density operator of the system that is obtained by tracing the total density operator over the bath in the interaction picture \cite{bre02,gar04,ali07,riv12}
\begin{equation}
	\frac{d\tilde{\rho}_\text{S}(t)}{dt}=-\frac{1}{\hbar^2}\int_0^t {d}s\Tr_\text{B}\comm{\tilde{H}_\text{I}(t)}{\comm{\tilde{H}_\text{I}(t-s)}{\tilde{\rho}(t)}},\!
	\label{eq:driven_Redfield_general}
\end{equation} 
with $\tilde{\rho}(t)=\tilde{\rho}_\text{S}(t)\otimes{\rho_\text{B}^\beta}$ inside the double commutator. An arbitrary operator $O$ has the interaction picture representation $\tilde{O}(t)=U^\dagger(t)OU(t)$ with the total time evolution operator $U(t)=U_\text{S}(t)U_\text{B}(t)$. The system unitary $U_\text{S}(t)$ is determined by ${i}\hbar dU_\text{S}(t)/dt=H_\text{S}(t)U_\text{S}(t)$ and the bath unitary reads $U_\text{B}(t)=\exp(- {i}H_\text{B}t/\hbar)$. Inserting the interaction Hamiltonian $H_I$ into Eq.~\eqref{eq:driven_Redfield_general} yields the equation
\begin{eqnarray}
		\frac{d\tilde{\rho}_\text{S}(t)}{dt}&=&\int_0^t {d}s\left(C_{12}(s)\comm{\tilde{a}(t-s)\tilde{\rho}_\text{S}(t)}{\tilde{a}^\dagger(t)}\right. \nonumber \\
		&+&\left.C_{21}(s)\comm{\tilde{a}^\dagger(t-s)\tilde{\rho}_\text{S}(t)}{\tilde{a}(t)}\right) + \text{h.c.},
	\label{eq:driven_Redfield_HO}
\end{eqnarray}
where we have defined the bath correlation functions, $C_{ij}(s)=\langle \tilde{B}_i(s)B_j\rangle_\beta/\hbar^2$, with the operators $B_1=b$, $B_2=b^\dagger$ and the ensemble average $\langle\cdot\rangle_\beta=\Tr_\text{B}(\cdot{\rho_\text{B}^\beta})$ \cite{bre02,gar04,ali07,riv12}; we have further used $C_{ii}(s)=0$ \cite{bre02,gar04,ali07,riv12}. Equation \eqref{eq:driven_Redfield_HO} has the form of a  driven Redfield master equation: it is similar to the coarse-grained master equation obtained in Ref.~\cite{moz20} for driven systems with a finite-dimensional Hilbert space. The Redfield master equation does not, in general,  preserve positivity \cite{bre02,gar04,ali07,riv12}. However, in our case, the interaction Hamiltonian $H_I$ conserves the total number of quanta, therefore  ensuring the positivity of Eq.~\eqref{eq:driven_Redfield_HO}.
 
The next step is to add the interaction picture representations of the ladder operators $a$ and $a^\dagger$ into Eq.~\eqref{eq:driven_Redfield_HO}. It is usually difficult to find closed form expressions of the latter, unless the Hamiltonian is spanned by the basis of a finite Lie algebra \cite{aga70,fer89}. In the present case,  the Lie algebra of the  oscillator is spanned by the operators $\mathds{1}$, $a$, $a^\dagger$ and $n$.  As a result, we find (\cref{app_sec:a_a_dagger_ip})
\begin{equation}
	\tilde{a}(t)={{e}^{-{i}\omega t}}(A(t)+a),
	\label{eq:a_tilde}
\end{equation} 
where the function $A(t)={({i}\omega/2)}\int_0^t {d}t^\prime\lambda(t^\prime)\exp( {i}\omega t)$ depends explicitly on the driving  parameter $\lambda(t)$. Using Eq.~\eqref{eq:a_tilde}, we eventually obtain the driven nonadiabatic quantum master equation in the interaction picture (with Liouvillian superoperator $\tilde{\mathcal{L}}_t$)
\begin{eqnarray}
	\frac{d\tilde{\rho}_\text{S}(t)}{dt}&=&{-\frac{ {i}}{\hbar}\comm{H_\text{LS}+\tilde{K}(t)}{\tilde{\rho}_\text{S}(t)}+\mathcal{D}\left[\tilde{\rho}_\text{S}(t)\right]} \nonumber\\
		&=&\tilde{\mathcal{L}}_t\left[\tilde{\rho}_\text{S}(t)\right],
	\label{eq:nonad_me_ip}
\end{eqnarray}
where have introduced the (often negligible) Lamb shift Hamiltonian, $H_\text{LS}=\hbar \Sigma_{12}(\omega)a^\dagger a+\hbar \Sigma_{21}(\omega)aa^\dagger$, and the amplitude damping dissipator $\mathcal{D}\left[\cdot\right]$
\begin{eqnarray}
		\mathcal{D}\left[\cdot\right]&=\gamma_{12}(\omega)\left(a\cdot a^\dagger-\frac{1}{2}\acomm{a^\dagger a}{\cdot}\right)\nonumber \\
		&+\gamma_{21}(\omega)\left(a^\dagger\cdot a-\frac{1}{2}\acomm{a a^\dagger}{\cdot}\right)
		\label{eq:diss}
\end{eqnarray}
{of the undriven system.} The damping coefficients are here given by $\gamma_{ij}(\omega)=\Gamma_{ij}(\omega)+\Gamma_{ij}^\ast(\omega)$ with $\Gamma_{ij}(\omega)=\int_0^\infty {d}s \,C_{ij}(s)\exp( {i}\sigma_{ij}\omega s)$ with $\sigma_{12}=-\sigma_{21}=1$, and $\Sigma_{ij}(\omega)=(1/2 {i})(\Gamma_{ij}(\omega)-\Gamma_{ij}^\ast(\omega))$. The effect of the external time-dependent driving is included in the term 
	$\tilde{K}(t)= {i}\hbar[f^\ast(t)a-f(t)a^\dagger]$,
with the driving function
\begin{equation}
	f(t)=\int_0^t {d}s\,{C(s)}
	{\mathrm{e}^{{i}\omega s}}A(t-s),
	\label{eq:driving_f}
\end{equation}
where we have defined  $C(s)=C_{12}(s)-C_{21}^\ast(s)$. Equation \eqref{eq:driving_f} vanishes in the absence of driving, $\lambda(t) =0$. In that case, Eq.~\eqref{eq:nonad_me_ip} reduces to the familiar amplitude damping master equation for the static harmonic oscillator \cite{bre02,gar04,ali07,riv12}.
We note that we have extended the upper integration limit of $\Gamma_{ij}(\omega)$ to infinity by assuming a fast decay of the bath correlation functions, as commonly done  \cite{bre02,gar04,ali07,riv12}. This is in general not possible with the upper integration limit of $f(t)$ because of the function $A(t-s)$, as this might lead to diverging integrals (see discussion below). 

In the continuum limit, the correlation functions $C_{ij}(s)$ can be expressed in terms of the spectral density $J(\omega)$ of the bath \cite{bre02,gar04,ali07,riv12}, see \cref{app_sec:damp_coeff_driv_func}. One obtains $\gamma_{12}(\omega)={2}\pi J(\omega)[n_\beta(\omega)+1]$ and $\gamma_{21}(\omega)={2}\pi J(\omega)n_\beta(\omega)$, with the thermal occupation number $n_\beta(\omega)=\left[\exp(\beta\hbar\omega)-1\right]^{-1}$. Furthermore, using the relation $\int_0^\infty {d}s{e}^{{i}xs}=\delta(x)+{i}\mathcal{P}\frac{1}{x}$, where $\mathcal{P}$ is the principal value \cite{kus06}, one finds  $\Sigma_{12}(\omega)=-\mathcal{P}\int_0^\infty{d}xJ(x)(n_\beta(x)+1)/(x-\omega)$ and  $\Sigma_{21}(\omega)=\mathcal{P}\int_0^\infty{d}xJ(x)n_\beta(x)/(x-\omega)$. Concretely choosing an Ohmic spectral density,  $J(\omega)=\eta\omega\exp(-\omega/\Omega)$, where $\eta$ is a dimensionless coupling constant and $\Omega$  is a cutoff frequency \cite{bre02,gar04,ali07,riv12}, {the function $C(s)$ in} the driving function \eqref{eq:driving_f} is explicitly given by (\cref{app_sec:damp_coeff_driv_func}) 
\begin{equation}
	{C(s)= \frac{\eta}{\left(\frac{1}{\Omega}+ {i} s\right)^2}.} 
	\label{eq:C(s)}
\end{equation}

Transforming back to the Schrödinger picture, the nonadiabatic master equation \eqref{eq:nonad_me_ip} becomes (\cref{app_sec:master_equation_Schroedinger})
\begin{eqnarray}
	\frac{d\rho_\text{S}(t)}{dt}&=&{-\frac{ {i}}{\hbar}\comm{H_\text{S}(t)+H_\text{LS}+K(t)}{\rho_\text{S}(t)}+\mathcal{D}\left[\rho_\text{S}(t)\right]}\nonumber\\&=&  {\mathcal{L}}_t\left[{\rho}_\text{S}(t)\right],
	\label{eq:nonad_me}
\end{eqnarray}
with  the operator $K(t)={i}\hbar[g^\ast(t)a-g(t)a^\dagger]$ and the driving function
\begin{equation}
	{g(t)=e^{-{i}\omega t}[f(t)-\alpha(\omega)A(t)].}
	\label{eq:driving_g}
\end{equation}
The constant $\alpha(\omega)$ is moreover given by $\alpha(\omega)=\gamma(\omega)/2+{i}\Sigma(\omega)$ where $\gamma(\omega)=\gamma_{12}(\omega)-\gamma_{21}(\omega)$ and $\Sigma(\omega)=\Sigma_{12}(\omega)+\Sigma_{21}(\omega)$. The effect of the driving in the Schrödinger picture is thus both encoded in the system Hamiltonian $H_\text{S}(t)$ \eqref{eq:H_S} and in the driving function $g(t)$ \eqref{eq:driving_g}. Note that, since the external driving is linear in the position of the harmonic oscillator, the dissipators remain here time independent. We furthermore emphasize that the Markovian master equation \eqref{eq:nonad_me} is of Lindblad form, meaning that its Liouvillian $\mathcal{L}_t$ generates a completely-positive trace preserving (CPTP) map, which guarantees that a proper density matrix is mapped onto a proper density matrix at all times \cite{bre02,gar04,ali07,riv12}.

\section{Recovering  the adiabatic  master equation}
\label{sec:ad_and_wd_me}

The nonadiabatic master equation \eqref{eq:nonad_me} is valid for arbitrary driving parameter $\lambda(t)$. In this section, we show how the adiabatic master equation may be recovered from it in the appropriate limit. A similar analysis for the weakly driven master equation \cite{ali79,ali06,gev95,koh99} is presented in \cref{app_sec:wd_me}. In the derivation of the master equation \eqref{eq:nonad_me}, the {interaction picture representation $\tilde{O}_\text{S}(t)$} of any system observable $O_\text{S}$ obeys the  exact equation $\tilde{O}_\text{S}(t)=U_\text{S}^\dagger(t)O_\text{S}U_\text{S}(t)$, where $U_\text{S}(t)$ is the  time evolution operator of the system. By contrast, for adiabatic driving, the system dynamics is approximated by  $\tilde{O}_\text{S}^\text{ad}(t)=U_\text{S}^{\text{ad}\dagger}(t)O_\text{S}U_\text{S}^\text{ad}(t)$, where the adiabatic time evolution operator $U_\text{S}^\text{ad}(t)$ is given by \cite{teu03}
\begin{equation}
	U_\text{S}^\text{ad}(t)=\sum_n\dyad{\varepsilon_n(t)}{\varepsilon_n(0)} {e}^{- {i}\mu_n(t)/\hbar},
	\label{eq:adiabatic_time_evolution}
\end{equation}
with the phase $\mu_n(t)=\int_0^t {d}\tau\left[\varepsilon_n(\tau)-\phi_n(\tau)\right]$ and the Berry connection $\phi_n(t)= {i}\hbar\braket{\varepsilon_n(t)}{\dot{\varepsilon}_n(t)}$; here, $\varepsilon_n(t)$ and $\ket{\varepsilon_n(t)}$ are the respective instantaneous eigenvalues and eigenstates of the {system} Hamiltonian $H_\text{S}(t)$. For two times, we furthermore have $\tilde{O}_\text{S}^\text{ad}(t-s)=\bar{U}_\text{S}^{\text{ad}\dagger}(t,s)O\bar{U}_\text{S}^\text{ad}(t,s)$ with  $\bar{U}_\text{S}^\text{ad}(t,s)=\exp({ {i}sH_\text{S}(t)/\hbar})U_\text{S}^\text{ad}(t)$. Repeating the above derivation  of the master equation for adiabatically evolving open quantum systems, we obtain (\cref{app_sec:details_ad_me}) 
\begin{eqnarray}
		\frac{d\tilde{\rho}_\text{S}(t)}{dt}&=&{-\frac{ {i}}{\hbar}\comm{H_\text{LS}+\tilde{K}_\text{ad}(t)}{\tilde{\rho}_\text{S}(t)}+\mathcal{D}\left[\tilde{\rho}_\text{S}(t)\right]}\nonumber \\
		&=&\tilde{\mathcal{L}}_{t,\text{ad}}\left[\tilde{\rho}_\text{S}(t)\right].
	\label{eq:ad_me_ip}
\end{eqnarray}
The adiabatic master equation \eqref{eq:ad_me_ip} is analogous to  Eq.~(47) of Ref.~\cite{alb12}. It has the same form as the nonadiabatic master equation \eqref{eq:nonad_me_ip} with the  difference   that the driving term $\tilde{K}(t)$ is  now replaced by the adiabatic approximation  $\tilde{K}_\text{ad}(t)= {i}\hbar[f^\ast_\text{ad}(t)a-f_\text{ad}(t)a^\dagger]$, where the adiabatic driving function $f_\text{ad}(t)$ reads
\begin{equation}
	f_\text{ad}(t)=\int_0^t {d}s\, {C(s)}{e}^{{i}\omega s}{A_\text{ad}(t,s)},
	\label{eq:driving_f_ad}
\end{equation}
{where we have introduced the function ${A_\text{ad}(t,s)}={[\lambda(t)/2]}\exp[ {i}\omega (t-s)]$.} 

We next express the adiabatic master equation \eqref{eq:ad_me_ip} in the Schrödinger picture, performing,  for consistency,  the transformation with $U_\text{S}^\text{ad}(t)$ instead of $U(t)$. We obtain
\begin{eqnarray}
	\frac{d\rho_\text{S}(t)}{dt}&=&-\frac{ {i}}{\hbar}\comm{H_\text{S}^\text{ad}(t)+H_\text{LS}+K_\text{ad}(t)}{\rho_\text{S}(t)}+\mathcal{D}\left[\rho_\text{S}(t)\right]\nonumber\\
	&=&{\mathcal{L}}_{t,\text{ad}}\left[{\rho}_\text{S}(t)\right],
	\label{eq:ad_me}
\end{eqnarray}
with the operator $K_\text{ad}(t)={i}\hbar[g_\text{ad}^\ast(t)a-g_\text{ad}(t)a^\dagger]$ and the adiabatic driving function
\begin{equation}
	g_\text{ad}(t)={e}^{-{i}\omega t}[f_\text{ad}(t)-\alpha(\omega)A_\text{ad}(t)]
	\label{eq:driving_g_ad}
\end{equation}
with $A_\text{ad}(t)=[\lambda(t)/2]\exp({i}\omega t)$. The adiabatic system Hamiltonian moreover reads $H_\text{S}^\text{ad}(t)=H_\text{S}(t)+{i}\hbar(\dot{\lambda}(t)/2)[a^\dagger-a]$. The effect of the adiabatic driving is hence encoded in the adiabatic Hamiltonian  $H_\text{S}^\text{ad}(t)$ and in the adiabatic driving function $g_\text{ad}(t)$ \eqref{eq:driving_g_ad}.

In order to  compare the nonadiabatic and adiabatic time-dependent master equations \eqref{eq:nonad_me} and \eqref{eq:ad_me}, it is convenient to split the function $A(t-s)$ in Eq.~\eqref{eq:driving_f} as 
\begin{equation}
	A(t-s)={A_\text{ad}(t,s)+\delta A_1(t,s)+\delta A_2(t,s)},
	\label{eq:separation_of_A}
\end{equation}
where, after partial integration, adding a zero {and using $\lambda(0)=0$}, we have identified the two nonadiabatic {contributions}
\begin{eqnarray}
	{\delta A_1(t,s)}&=&\frac{\lambda(t-s)-\lambda(t)}{2}{{e}^{ {i}\omega(t-s)}},\\
	{\delta A_2(t,s)}&=&-\int_0^{t-s} {d}t^\prime\frac{\dot{\lambda}(t^\prime)}{2}{{e}^{ {i}\omega t^\prime}}.
	\label{eq:parts_of_A}%
\end{eqnarray}%
Evaluating Eq.~\eqref{eq:separation_of_A} at $s=0$ yields $A(t)=A_\text{ad}(t)+\delta A(t)$ with $\delta A(t)=\delta A_2(t,0)$. Further inserting the  decomposition \eqref{eq:separation_of_A} into Eq.~\eqref{eq:driving_f}, we  obtain $f(t)=f_\text{ad}(t)+\delta f(t)$, where the nonadiabatic contribution $\delta f(t)$ to the driving function  is given by
\begin{equation}
	\delta f(t)=\int_0^t {d}s\,{C(s)}{e}^{{i}\omega s}\left[{\delta A_1(t,s)+\delta A_2(t,s)}\right].
	\label{driving_delta_f}
\end{equation}
{We can then  write $g(t)=g_\text{ad}(t)+\delta g(t)$ with the nonadiabatic contribution
\begin{equation}
	\delta g(t)={e}^{-{i}\omega t}[\delta f(t)-\alpha(\omega)\delta A(t)].
	\label{eq:delta_g}
\end{equation}
The nonadiabatic master equation \eqref{eq:nonad_me} finally  reads
\begin{equation}
	\frac{d{\rho_\text{S}(t)}}{dt}=\mathcal{L}_{t,\text{ad}}\left[\rho_\text{S}(t)\right]-\frac{ {i}}{\hbar}\comm{\delta H_\text{S}(t)+\delta K(t)}{\tilde{\rho}_\text{S}(t)},
	\label{eq:nonad_me_vs_ad_me}
\end{equation}
where  $\mathcal{L}_{t,\text{ad}}$ is the Liouvillian of the adiabatic master equation \eqref{eq:ad_me}, and the operators $\delta H_\text{S}(t)=H_\text{S}(t)-H_\text{S}^\text{ad}(t)$ and $\delta K(t)= {i}\hbar[\delta g^\ast(t)a-\delta g(t)a^\dagger]$ determine the nonadiabatic correction.

Equations \eqref{eq:separation_of_A}-\eqref{eq:nonad_me_vs_ad_me} provide an intuitive understanding of the approach to the adiabatic limit. For slow adiabatic driving, the time derivative $\dot{\lambda}(t)$ {becomes small, and thus $\delta H_\text{S}(t)$ as well  as $\delta A_2(t,s)$ and $\delta A(t)$ converge to zero. Moreover,} the difference $\lambda(t-s)-\lambda(t)$ in {$\delta A_1(t,s)$} {is small}, if $s$ is much smaller than the timescale on which $\lambda(t)$ changes significantly. Moreover, due the prefactor $C(s)$, the integrand in Eq.~\eqref{driving_delta_f} only gives relevant contributions to the integral for small $s$, rendering $\delta f(t)$ negligible. As a result, $\delta g(t)$, and hence the whole nonadiabatic contribution $\delta K(t)$, vanish in that limit. This argument is made more rigorous for the concrete case of a linear ramp in the next section.  

\section{Linear driving protocol}
\label{sec:ramp}

For concreteness, let us now  assume that the driving protocol is a linear ramp,  $\lambda(t)=\Delta l\,t/T$, where $\Delta l$ describes the displacement of the harmonic potential in units of $x_0$ during the total driving time $T$. The driving is intuitively fast for small $T$ and slow for large $T$. It is advantageous to use dimensionless quantities to simplify the analysis. We thus define the driving time in units of the system timescale, $\mathcal{T}=\omega T$, and the cutoff frequency in units of the system frequency, $w=\Omega/\omega$.  In this case, the function $A(t)$, Eq.~\eqref{eq:a_tilde},  is given by  $A(t)=A_\text{ad}(t)+\delta A(t)$ with the adiabatic part $A_\text{ad}(t)=(\Delta l/2)(t/T)\exp(i\omega t)$ and the nonadiabatic contribution $\delta A(t)=(\Delta l/2)({i}/\omega T)(\exp({i}\omega t)-1)$. The nonadiabatic driving function $f(t)$ can  accordingly be written in the form
\begin{equation}
	f(t)=\frac{\eta\Delta l}{{2}T}\left[F_\text{ad}\left(\frac{t}{T}\right)+\delta F\left(\frac{t}{T}\right)\right],
	\label{eq:driving_f_3}
\end{equation}
where we have introduced $f_\text{ad}(t)=(\eta\Delta l/{2}T)F_\text{ad}(t/T)$ and $\delta f(t)=(\eta\Delta l/{2}T)\delta F(t/T)$. The two functions $F_\text{ad}(\tau)$ and $\delta F(\tau)$ are additionally given by 
\begin{subequations}
	\begin{align}
		F_\text{ad}(\tau)&=w\mathcal{T}\tau\exp( {i}\mathcal{T}\tau)I_0(w\mathcal{T}\tau),\\
		\delta F(\tau)&= {e}^{ {i}\mathcal{T}\tau}\left[ {i}wI_0(w\mathcal{T}\tau)-I_1(w\mathcal{T}\tau)\right]\nonumber\\
		&- {i}wI_\text{e}(w\mathcal{T}\tau),
	\end{align}
	\label{eq:part_of_f}%
\end{subequations}%
with the dimensionless time $\tau=t/T$ and the integrals
\begin{subequations}
	\begin{align}
		I_0(z)&=\int_0^{z} {d}x\frac{1}{(1+ {i}x)^2},\\
		I_1(z)&=\int_0^{z} {d}x\frac{x}{(1+ {i}x)^2},\\
		I_\text{e}(z)&=\int_0^{z} {d}x\frac{ {e}^{ {i}x/w}}{(1+ {i}x)^2}.
	\end{align}
	\label{eq:integrals}%
\end{subequations}%
Exact analytic expressions for these three integrals as well as their asymptotic behavior in the limit of large $z$ can be found in \cref{app_sec:integrals}. We can then construct $g(t)$, $g_\text{ad}(t)$ and $\delta g(t)$ from Eqs.~\eqref{eq:driving_g}, \eqref{eq:driving_g_ad} and \eqref{eq:delta_g}.

\begin{figure}[t]
	\includegraphics[width=\columnwidth]{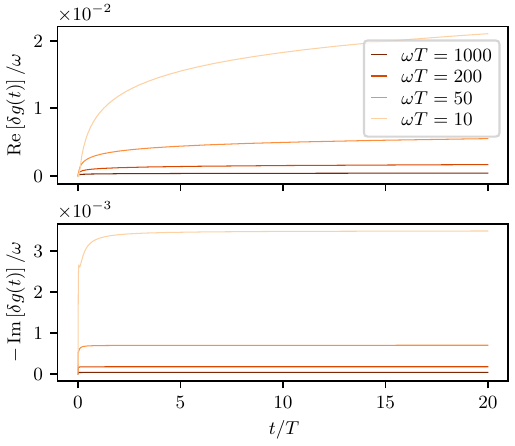}
	\caption{Nonadiabatic contribution to the driving function. The figure displays the real part (first row) and minus the imaginary part (second row) of the nonadiabatic contribution $\delta g(t)$, Eq.~\eqref{eq:delta_g}, of the driving function $g(t)$, Eq.~\eqref{eq:driving_g}, in units of system frequency $\omega$ as a function of the dimensionless time $t/T$, for different driving speeds characterized by $\omega T$. Other parameters are   $\Omega/\omega=4$, $\eta=0.008$ and $\Delta l=10.$}
	\label{fig:delta_g}
\end{figure}

Using the above expressions, the approach to the adiabatic limit may be formulated in a more precise way. According to Eq.~\eqref{eq:nonad_me_vs_ad_me}, the nonadiabatic master equation \eqref{eq:nonad_me_ip} reduces to the adiabatic master equation \eqref{eq:ad_me_ip} when {$\delta H_\text{S}(t)$ and  $\delta K(t)$} tend to zero. We expect this to happen in the limit of long driving times, $T\to\infty$. However, to properly define the adiabatic limit, one cannot only naively let $T$ run to infinity. One has to  additionally  keep the dimensionless time $\tau= t/T$ constant. In that limit, the functions $\delta A(t)$ and $\delta F(t/T)/T$ in Eq.~\eqref{eq:driving_f_3}, and by that also $\delta g(t)$ as well as $\dot{\lambda}(t)$, indeed converge to zero, implying that 
$g(t) \rightarrow g_\text{ad}(t)$. As a consequence
\begin{equation}
	{\lim_{\substack{T\to\infty\\\frac{t}{T}\;\text{const}}}\delta H_\text{S}(t)=}\lim_{\substack{T\to\infty\\\frac{t}{T}\;\text{const}}}{\delta K(t)}=0.
	\label{eq:ad_limit}
\end{equation}
According to the above equations, the nonadiabatic contribution to the quantum master equation vanishes for infinitely slow driving. \Cref{fig:delta_g} displays the real part (first row) and minus the imaginary part (second row) of the nonadiabatic contribution $\delta g(t)$ defined in Eq.~\eqref{eq:delta_g} of the driving function $g(t)$ defined in Eq.~\eqref{eq:driving_g} in units of system frequency $\omega$ as function of the dimensionless time $t/T$, for different driving speeds characterized by $\omega T$. One can indeed observe that it vanishes when the external driving becomes adiabatically slow.

\section{Solving the master equation}
\label{sec:adjoint_me_and_obs}

We proceed by determining the time evolution of the expectation value, $\expval{O}\!(t)=\Tr(O\rho_\text{S}(t))$,  of an observable $O$ by solving the corresponding adjoint master equation \cite{bre02,gar04} (the explicit solution for the density operator $\rho_\text{S}(t)$ is provided in \cref{app_sec:solve_me}). The Liouvillians of the nonadiabatic and adiabatic master equations \eqref{eq:nonad_me} and \eqref{eq:ad_me}  generate Gaussian preserving maps \cite{ser17}, implying that if the  state is initially Gaussian, it will remain Gaussian at all times. We will thus limit our analysis to Gaussian initial states. An important property of a  master equation generating a Gaussian preserving map is that the adjoint master equations for the first and second moments of its solution are decoupled from higher moments.  Starting with the nonadiabatic master equation \eqref{eq:nonad_me}, one finds that the adjoint master equation for $\expval{O}\!(t)$ reads \cite{bre02,gar04}
\begin{equation}
	\dv{t}\expval{O}\!(t)=\expval{\mathcal{L}_t^\star\left[O\right]}\!(t).
	\label{eq:adj_me_gen}
\end{equation} 
The dual Liouvillian $\mathcal{L}_t^\star$ is here given by
\begin{equation}
	\mathcal{L}_t^\star\left[\cdot\right]=	\frac{\mathrm{i}}{\hbar}\comm{H_\text{S}(t)+H_\text{LS}+K(t)}{\cdot}+\mathcal{D}^\star\left[\cdot\right],
	\label{eq:dual_Liouvillian}
\end{equation}
where we have defined the dual dissipator
\begin{equation}
	\begin{aligned}
		\mathcal{D}^\star\left[\cdot\right]&=\gamma_{12}(\omega)\left(a^\dagger\cdot a-\frac{1}{2}\acomm{a^\dagger a}{\cdot}\right)\\
		&+\gamma_{21}(\omega)\left(a\cdot a^\dagger-\frac{1}{2}\acomm{a a^\dagger}{\cdot}\right).
		\label{eq:dual_diss}
	\end{aligned}
\end{equation}
Choosing for $O$ the operators $a$, $a^\smallsquare$ and $n$, and introducing the variance $V_a(t)=\expval{a^\smallsquare}\!(t)-\expval{a}^\smallsquare\!(t)$ and the covariance $C_{aa^{\text{\raisebox{-2pt}{$\dagger$}}}}(t)=\expval{n}\!(t)-\expval{a^{\text{\raisebox{-2pt}{$\dagger$}}}}\!(t)\!\expval{a}\!(t)$, we obtain
\begin{subequations}
	\begin{align}
		\dv{t}\!\expval{a}\!(t)&=-\delta(\omega)\!\expval{a}\!(t)-h(t),\\
		\dv{t}V_a(t)&=-2\delta(\omega)V_a(t),\\
		\dv{t}C_{aa^\smalldagger}(t)&=\gamma(\omega)\left[n_\beta(\omega)-C_{aa^\smalldagger}(t)\right],
	\end{align}
	\label{eq:adj_me_cm}%
\end{subequations}%
with the coefficient $\delta(\omega)=\alpha(\omega)+\mathrm{i}\omega$ and the function $h(t)=g(t)-({\mathrm{i}\omega}/{2})\lambda(t)$. In the case of the adiabatic master equation \eqref{eq:ad_me}, we respectively replace $H_\text{S}(t)$ and $K(t)$ by $H_\text{S}^\text{ad}(t)$ and $K_\text{ad}(t)$ in Eq.~\eqref{eq:dual_Liouvillian}, so that the function $h(t)$ is replaced by  $h_\text{ad}(t)=g_\text{ad}(t)-({\mathrm{i}\omega}/{2})\lambda(t)-{\dot{\lambda}(t)}/{2}$.
	
The choice of $\expval{a}\!(t)$ as  first moment and of $V_a(t)$ and $C_{aa^\smalldagger}(t)$ as second order moments is  advantageous, since the dynamics of all three quantities  decouple, and can hence be easily solved analytically. Furthermore, the second order moments are independent of the driving. They are the same for the two driven master equations:
\begin{subequations}
	\begin{align}
		V_a(t)&=V_a(0){e}^{-2\delta(\omega)t},\\
		C_{aa^{\text{\raisebox{-2pt}{$\dagger$}}}}(t)&=C_{aa^{\text{\raisebox{-2pt}{$\dagger$}}}}(0){e}^{-\gamma(\omega)t} + n_\beta(\omega)\left(1-{e}^{-\gamma(\omega)t}\right).
		\label{eq:V_and_C}
	\end{align}
\end{subequations}
This implies that $V_a(t)$ and $C_{aa^{\text{\raisebox{-2pt}{$\dagger$}}}}(t)$ converge to their equilibrium values set by the  inverse bath temperature $\beta$. On the other hand, the first moment $\expval{a}\!(t)$ explicitly depends on the driving via the function $h(t)$, and reads
\begin{equation}
	\expval{a}\!(t)=\expval{a}\!(0){e}^{-\delta(\omega)t} - {e}^{-\delta(\omega)t}\int_0^t{d}t^\prime{e}^{\delta(\omega)t^\prime}h(t^\prime).
	\label{eq:exp_val_a}
\end{equation}
For the adiabatic master equation \eqref{eq:ad_me}, the function $h(t)$ is again replaced by $h_\text{ad}(t)$. 

\begin{figure*}[htb]
	\includegraphics[width=\textwidth]{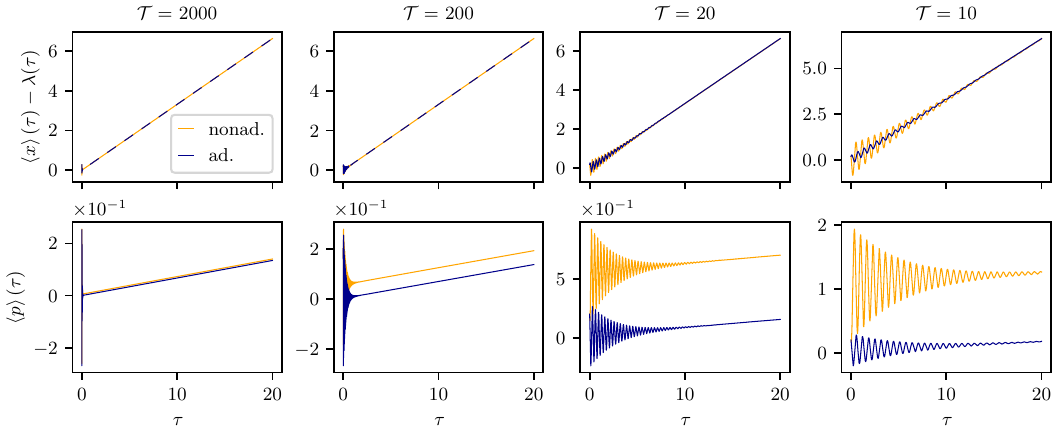}
	\caption{Position and momentum. Expectation value of the dimensionless position to the minimum of the harmonic potential $\expval{x}\!(\tau)-\lambda(\tau)$ (upper row) as well as  expectation value of the dimensionless momentum $\expval{p}\!(\tau)$ (lower row) as a function of the dimensionless time $\tau$ for  the nonadiabatic master equation \eqref{eq:nonad_me} (orange) and the adiabatic master equation \eqref{eq:ad_me} (blue) for different driving speeds: very slow ($\mathcal{T}=2000$, first column), slow ($\mathcal{T}=200$, second column), intermediate ($\mathcal{T}=20$, third column) and fast driving ($\mathcal{T}=10$, fourth column). The other parameters are fixed at the values $y=0.1$, $w=4$, $\eta=0.008$, $\Delta l=10$ and the initial state is determined by $\expval{a}(0)=0.1+0.1{i}$, $V(0)=0$ and $C(0)=n_\text{th}+\delta n_0-|\expval{a}(0)|^{\raisebox{-2pt}{\scriptsize 2}}$ with $\delta n_0=2$.}
	\label{fig:x_and_p}%
\end{figure*}

The numerical evaluation of the above moments is facilitated by expressing them in dimensionless form. Introducing the dimensionless time $\tau=t/T$, we have
\begin{subequations}
	\begin{align}
		\expval{a}\!(\tau)&=\expval{a}\!(0){e}^{-\bar{\delta}\tau}-{e}^{-\bar{\delta}\tau}\int_0^\tau{d}\tau^\prime{e}^{\bar{\delta}\tau^\prime}\bar{h}(\tau^\prime),\\
		V_a(\tau)&=V_a(0){e}^{-2\bar{\delta}\tau},\\
		C_{aa^\smalldagger}(\tau)&=C_{aa^\smalldagger}(0){e}^{-\bar{\gamma}\tau}+n_\text{th}\left(1-{e}^{-\bar{\gamma}\tau}\right),
	\end{align}
	\label{eq:a_and_V_and_C_dimensionless}%
\end{subequations}%
where we rewrote the thermal occupation number as $n_\text{th}=\left[\exp(1/y)-1\right]^{-1}$, with $y=1/(\beta\hbar\omega)$ the thermal energy in units of the system energy. We have additionally defined the dimensionless constants $\bar{\delta}=T\delta(\omega)$ and $\bar{\gamma}=T\gamma(\omega)$, as well as the dimensionless driving function $\bar{h}(\tau)=Th(T\tau)$. We analogously define $\bar{h}_\text{ad}(\tau)=Th_\text{ad}(T\tau)$ for the adiabatic master equation \eqref{eq:ad_me}. Detailed expressions for the dimensionless quantities can be found in \cref{app_sec:dimless}. We emphasize that Eq.~\eqref{eq:a_and_V_and_C_dimensionless}, and thereby the solution of the nonadiabatic master equation, is fully characterized by the five dimensionless parameters $y$, $w$, $\eta$, $\mathcal{T}$ and $\Delta l$.

\section{Position and momentum}
\label{sec:pm}

In the following, we compare the predictions of the nonadiabatic master equation \eqref{eq:nonad_me} and of the adiabatic master equation \eqref{eq:ad_me} for different observables and for various driving speeds for the linear ramp. We first focus on dynamical variables, and consider the dimensionless position and momentum operators, $x=(a^\smalldagger+a)$ and $p=\mathrm{i}(a^\smalldagger-a)$. The corresponding mean values, variances and covariances are respectively given by  $\expval{x}=2\Re(\expval{a})$ and $\expval{p}=2\Im(\expval{a})$, $V_x=\expval{x^\smallsquare}-\expval{x}^\smallsquare=2C_{aa^{\text{\raisebox{-2pt}{$\dagger$}}}} + 1 + 2\Re(V_a)$ and $V_p=\expval{p^\smallsquare}-\expval{p}^\smallsquare=2C_{aa^{\text{\raisebox{-2pt}{$\dagger$}}}} + 1 - 2\Re(V_a)$, as well as  $C_{xp}=(1/2)\expval{\acomm{x}{p}}-\expval{x}\!\!\expval{p}=2\Im(V_a)$. Again, only the first order moments depend on the driving.

\Cref{fig:x_and_p} shows the difference of the expectation value of the dimensionless position to the minimum of the harmonic potential $\expval{x}(\tau)-\lambda(\tau)$ (upper row) as well as the expectation value of the dimensionless momentum $\expval{p}(\tau)$ (lower row) as functions of the dimensionless time $\tau$ for the nonadiabatic master equation \eqref{eq:nonad_me} (orange lines) and the adiabatic master equation \eqref{eq:ad_me} (blue lines). Different driving speeds are considered: very slow driving with $\mathcal{T}=2000$ (first column), slow driving  with $\mathcal{T}=200$ (second column), intermediate driving with $\mathcal{T}=20$ (third column) and quick driving with $\mathcal{T}=10$ (fourth column). The other parameters are fixed at $y=0.1$, $w=4$, $\eta=0.008$ and $\Delta l=10$. {The initial state is chosen such that} $\expval{a}(0)=0.1+0.1{i}$, $V_a(0)=0$ and $C_{aa^\smalldagger}(0)=n_\text{th}+\delta n_0-|\expval{a}(0)|^{\raisebox{-2pt}{\scriptsize 2}}$ with $\delta n_0=2$. For both master equations and all driving speeds, we observe a drift away from the minimum of the harmonic potential that increases linearly in time and is superposed by decaying oscillations. The same holds true for the momentum, although we only observe a very slight linear increase. As anticipated, we observe almost perfect agreement between nonadiabatic and adiabatic master equation for very slow driving. For increasing driving speed, i.e. decreasing duration of the process, the predictions made by the two master equations deviate more and more. While we have only a difference in the oscillation amplitude that grows slightly with the driving speed  for the position, we find for the momentum an additional offset between the two asymptotically linear functions that strongly augments with the driving speed. These deviations stem from the differences between the  two functions $g(t)$ and $g_{ad}(t)$ displayed in \cref{fig:delta_g}, as well as in the difference between $H_\text{S}(t)$ and $H_\text{S}^\text{ad}(t)$. 

\section{Energy and entropy}
\label{sec:E_and_S}

We next analyze the behavior of thermodynamic quantities such as average energy and entropy of the system. Using Eq.~\eqref{eq:H_S}, the mean energy can be written  in units of $\hbar\omega$ and as function of the dimensionless time $\tau$ as
\begin{equation}
	 E(\tau)=\expval{n}\!(\tau)+\frac{1}{2}-\frac{\Delta l\tau}{2} 2\Re\expval{a}\!(\tau)+\frac{\Delta l^2\tau^2}{4}.
	 \label{eq:mean_energy_1}	
\end{equation}
It is useful to express it as a function of real first and second order moments of the state, which yields
\begin{eqnarray}
		E(\tau)&=&\frac{1}{4}\left(\expval{p}^2\!(\tau)+\expval{x}^2\!(\tau)\right)-\frac{\Delta l \tau}{2}\expval{x}\!(\tau)+\frac{\Delta l^2\tau^2}{4}\nonumber\\
		&+&\frac{1}{4}\left(V_p(\tau)+V_x(\tau)\right).
		\label{eq:mean_energy_2}
\end{eqnarray}
The main advantage of this form is that it clearly distinguishes between contributions that depend on the driving in the first line and contributions that are independent of the driving in the second line. \Cref{fig:energies} shows the mean energy $E(\tau)$ in units of $\hbar\omega$ as a function of the dimensionless time $\tau$ for the nonadiabatic master equation \eqref{eq:nonad_me} and the adiabatic master equation \eqref{eq:ad_me}. The color scheme, the initial state and the parameters $y$, $w$, $\eta$ and $\Delta l$ are the same as in \cref{fig:x_and_p}. We, however, chose different driving speeds ($\mathcal{T}\in\lbrace100,20,10,5\rbrace$), as deviations between the two master equations only become significant for much faster driving speeds than this was the case for  $\expval{x}\!(\tau)$ and $\expval{p}\!(\tau)$. For both master equations, the average energy initially decreases, which can be explained by the initial decay of the variances and the decaying components of position and momentum and their squares, while for longer times the nearly quadratic increase of squared position, momentum and driving protocol dominates. For fast driving, the mean energy furthermore exhibits oscillations. We note that both master equations perfectly agree for slow driving, whereas the adiabatic master equation strongly underestimates the amplitude of these oscillations for fast driving.

\begin{figure*}[htb]
	\includegraphics[width=\textwidth]{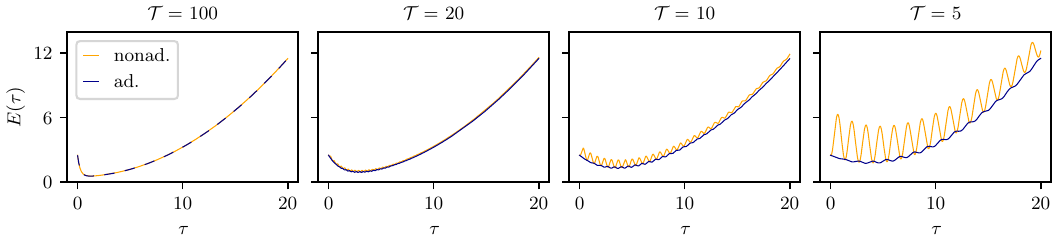}
	\caption{Energy. Mean value of the energy of the harmonic oscillator, in units of $\hbar\omega$, as a function of the dimensionless time $\tau=t/T$ for the nonadiabatic master equation \eqref{eq:nonad_me} (orange) and the adiabatic master equation \eqref{eq:ad_me} (blue) for different driving times $\mathcal{T}\in\lbrace100,20,10,5\rbrace$ from slow driving to fast driving. Same parameters as in \cref{fig:x_and_p}.}
	\label{fig:energies}%
\end{figure*}%

On the other hand, the von Neumann entropy, $S_\text{vN}(\rho)= - \Tr[{\rho}\ln {\rho}]$, for  a single mode Gaussian state can be evaluated, as a function of the dimensionless time $\tau$, as  \cite{ser17}
\begin{equation}
	S_\text{vN}(\tau)=s\left(\frac{\mu(\tau)+1}{2}\right)-s\left(\frac{\mu(\tau)-1}{2}\right),
	\label{eq:S_vN}
\end{equation}
with  the function $s(x)=x\ln(x)$ and $\mu(\tau)=[V_x(\tau)V_p(\tau)-C_{xp}^2(\tau)]^{-1/2}$. Since it only depends on second order moments of the Gaussian state, it is unaffected by the external driving. It is therefore identical for the two quantum master equations, as seen in \cref{fig:entropy} for the same initial state and the same values of $y$, $w$, $\eta$ as in \cref{fig:x_and_p}. It is, in particular, independent of $\Delta l$ and $\mathcal{T}$. The value of $\mathcal{T}$ only determines the transient time interval that we consider; here we set $\mathcal{T}=20$. The entropy quickly converges to a constant value, which corresponds to the thermal state with  inverse bath temperature $\beta$ and the initial system Hamiltonian $H_\text{S}(0)$. 

\begin{figure}[b]
	\includegraphics[width=\columnwidth]{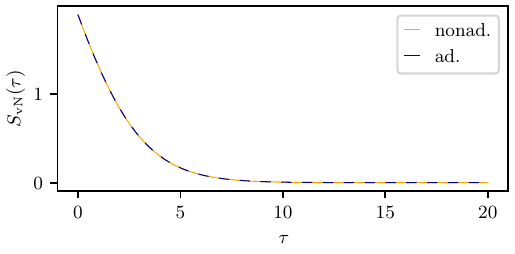}
	\caption{Entropy. Von Neumann entropy of the harmonic oscillator as a function of the dimensionless time $\tau=t/T$. The von Neumann entropy is independent of the driving and, thus, the same for the two driven master equations \eqref{eq:nonad_me} and \eqref{eq:ad_me} ($\mathcal{T}=20$ here). Same parameters as in \cref{fig:x_and_p}.}
	\label{fig:entropy}
\end{figure}

\section{Quantum coherence}
\label{sec:coherence}

The dissipative dynamics of an open quantum system does not only affect the diagonal density matrix elements (related to the energy), but also the nondiagonal density matrix elements (connected to quantum coherence) \cite{bre02,gar04,ali07,riv12}. In this section, we analyze the influence of the external driving on the quantum coherence \cite{str17} of the harmonic oscillator in the instantaneous energy eigenbasis.

\begin{figure*}[ht]
	\includegraphics[width=\textwidth]{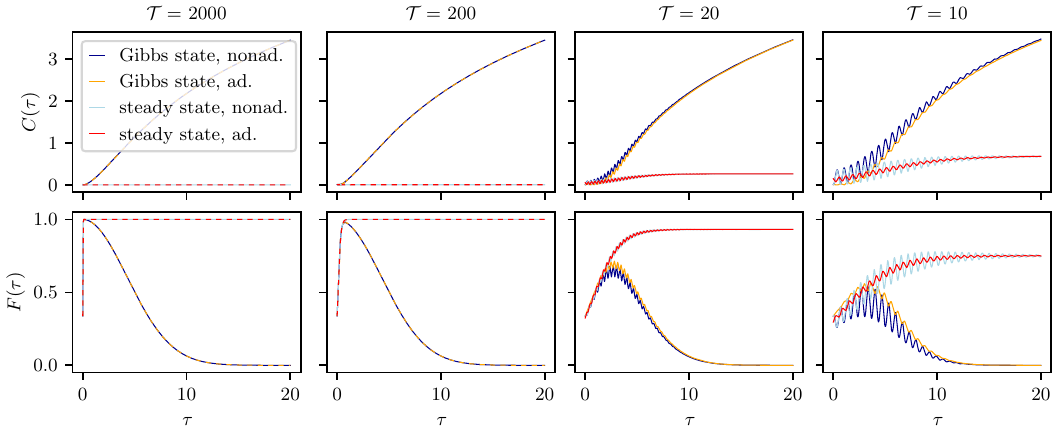}
	\caption{Quantum coherence and instantaneous steady state. The first row displays the  coherence measure $C_{\mathcal{B}_t}$ with respect to the instantaneous energy eigenbasis, Eq.~\eqref{eq:C_var_basis} (blue and orange lines), and with respect to the eigenbasis of the instantaneous steady state (light blue and red lines), as a function of the dimensionless time $\tau=t/T$. The second row displays the fidelity $\mathcal{F}$ between the system state and the instantaneous Gibbs state, Eq.~\eqref{41} (blue and orange lines), and the instantaneous steady state (light blue and red lines), as a function of the dimensionless time $\tau=t/T$. The results predicted by the nonadiabatic master equation \eqref{eq:nonad_me} (blue/ light blue) and the adiabatic master equation \eqref{eq:ad_me} (orange/ red) are shown for different driving speeds: very slow ($\mathcal{T}=2000$, first column), slow ($\mathcal{T}=200$, second column), medium ($\mathcal{T}=20$, third column) and fast driving ($\mathcal{T}=10$, fourth column). Same parameters as in \cref{fig:x_and_p}.}
	\label{fig:coherences}
\end{figure*}

A commonly used measure of quantum coherence of a state $\rho$ in a given basis is the relative entropy of coherence, $C_\text{rel}(\rho) =  S_\text{vN}(\rho_\text{d}) - S_\text{vN}(\rho)$, where $\rho_\text{d}$ is the diagonal part of $\rho$ in the chosen basis \cite{bau14}. This coherence monotone was originally introduced for finite-dimensional quantum systems \cite{bau14}. In the following, we use a related coherence quantifier for infinite-dimensional Gaussian states put forward in Ref.~\cite{xu16}. Considering a (fixed)  orthonormal basis $\mathcal{B}=\lbrace\ket{n}\rbrace_{n=0}^\infty$, a measure of the coherence of a state $\rho$ in that basis may be defined as \cite{xu16}
\begin{equation}
	C_\mathcal{B}(\rho)=\inf_\sigma\left\lbrace S(\rho||\sigma),\sigma\in\mathcal{I}\right\rbrace,
	\label{eq:C_initial_def}
\end{equation}
where $S(\rho||\sigma)=\Tr(\rho\ln\rho -\rho\ln\sigma)$ is the quantum relative entropy of $\rho$ with respect to $\sigma$ \cite{bre02,gar04,ali07} and $\mathcal{I}$ is the set of Gaussian states that are incoherent in $\mathcal{B}$. The quantum relative entropy is nonnegative, and vanishes only if the two states are equal. Although not a true distance measure, it characterizes the difference between two density operators \cite{bre02,gar04,ali07}. By taking the infimum over all $\sigma\in\mathcal{I}$, the coherence measure \eqref{eq:C_initial_def} quantifies the separation between $\rho$ and the closest incoherent Gaussian state. Any such (single mode) Gaussian state that is incoherent in $\mathcal{B}$ may be written as \cite{xu16}
\begin{equation}
	\sigma_\mathcal{B}(\bar{n})=\sum_{k=0}^\infty\frac{\bar{n}^k}{(\bar{n}+1)^{k+1}}\dyad{k}{k},
	\label{eq:inc_Gauss}
\end{equation}
where $\bar{n}=\Tr[n\sigma_\mathcal{B}(\bar{n})]$ is a nonnegative real number and $n$ is the number operator corresponding to the basis $\mathcal{B}$. The state $\sigma_\mathcal{B}(\bar{n})$ can be regarded as a thermal state of the Hamiltonian $H=\hbar{\omega}(n+1/2)$, with inverse temperature $\beta$, which is connected to $\bar{n}$ via $\bar{n}=[\exp(\beta\hbar\omega)-1]^{-1}$. Inserting Eq.~\eqref{eq:inc_Gauss} into Eq.~\eqref{eq:C_initial_def}, one finds
\begin{equation}
	C_\mathcal{B}(\rho)=-\max_{\bar{n}}\Tr[\rho\ln\sigma_\mathcal{B}(\bar{n})] -S_\text{vN}(\rho),
	\label{eq:C_def_with_max}
\end{equation}
where $S(\rho)$  is the von Neumann entropy of state $\rho$. The maximization over $\bar{n}$ can  be performed by setting the derivative of $\Tr[\rho\ln\sigma_\mathcal{B}(\bar{n})]$ with respect to $\bar{n}$ to zero, solving for $\bar{n}$, and inserting the corresponding result into Eq.~\eqref{eq:C_def_with_max} \cite{xu16}. One then obtains $\bar{n}_\text{max}=\langle n\rangle=\Tr[n\rho]$ and 
\begin{equation}
	\begin{aligned}
		C_\mathcal{B}(\rho)&=\left[\left(\langle n\rangle+1\right)\ln(\langle n\rangle+1)-\langle n\rangle\ln\langle n\rangle\right]\\
		&-S_\text{vN}(\rho).
	\end{aligned}
	\label{eq:C_const_basis}
\end{equation}
Equation \eqref{eq:C_const_basis} has  a simple interpretation: it is the quantum relative entropy of $\rho$ with respect to  the incoherent Gaussian state $\sigma_\mathcal{B}(\bar{n})$ with the same mean number $\bar n$.

For our purpose, we choose $\mathcal{B}_t=\lbrace\ket{\varepsilon_n(t)}\rbrace_{n=0}^\infty$ as the (time-dependent) instantaneous eigenbasis of the system Hamiltonian $H_\text{S}(t)$, and $\mathcal{I}_t$ as the set of Gaussian states that are incoherent in that basis. In analogy to Eq.~\eqref{eq:inc_Gauss}, any such incoherent state may be written as
\begin{equation}
	\sigma_{\mathcal{B}_t}(\bar{n}_t)=\sum_{k=0}^\infty\frac{\bar{n}_t^k}{(\bar{n}_t+1)^{k+1}}\dyad{\varepsilon_k(t)}{\varepsilon_k(t)},
\end{equation}
where, due to the external driving, $\bar{n}_t=\Tr[n_t\sigma_{\mathcal{B}_t}(\bar{n}_t)]= \Tr[n_t\rho_\text{S}(t)]$ is no longer the expectation value of $n$, but of  the  number operator corresponding to the basis $\mathcal{B}_t$
\begin{equation}
	\begin{aligned}
		n_t&=D\left({\frac{\lambda(t)}{2}}\right)nD^\dagger\left({\frac{\lambda(t)}{2}}\right)\\
		&=n-{\frac{\lambda(t)}{2}}\left(a^\dagger+a\right)+{\frac{\lambda^2(t)}{4}},
	\end{aligned}
\end{equation}
where $D$ is the displacement operator \cite{bre02,gar04}.
Equation \eqref{eq:C_const_basis} is accordingly replaced with
\begin{equation}
	\begin{aligned}
		C_{\mathcal{B}_t}(\rho_\text{S}(t))&=\left[\langle n_t\rangle(t)+1\right]\ln[\langle n_t\rangle(t)+1]\\
		&-\langle n_t\rangle(t)\ln[\langle n_t\rangle(t)]-S_\text{vN}(\rho_\text{S}(t)).
	\end{aligned}
	\label{eq:C_var_basis}
\end{equation}
In the first row of \cref{fig:coherences}, the blue and the orange lines show the coherence measure $C_{\mathcal{B}_t}$ with respect to the instantaneous energy eigenbasis, Eq.~\eqref{eq:C_var_basis}, for the driven oscillator as a function of the dimensionless time, $\tau=t/T$, for the same four driving times ($\mathcal{T}=2000,\,200,\,20,\,10$) as used previously in \cref{fig:x_and_p}. We note that {both} master equations yield different results for fast driving,  a behavior already seen  for the mean energy \eqref{eq:mean_energy_2} in \cref{sec:E_and_S}. In particular, the  driving leads to an increase of quantum coherence in the case of the adiabatic and nonadiabatic master equations \eqref{eq:ad_me} and \eqref{eq:nonad_me}, with a more pronounced enhancement for the nonadiabatic equation. 

\section{Instantaneous steady state}
\label{sec:iss}

In order to shed light on the features of the quantum coherence discussed in the above section, we next investigate the approach of the harmonic oscillator to the instantaneous Gibbs state. One intuitively expects that if a system weakly coupled to a heat bath is initially prepared in a thermal state and slowly driven on a timescale much longer than the relaxation timescale, it will stay close to the instantaneous Gibbs state. For a slowly driven nonthermal  initial state, one further expects the system to first relax to the instantaneous Gibbs state, and subsequently remain close to it. Contrary to this expectation, we shall show that the instantaneous steady states of the master equations do not coincide with the corresponding instantaneous Gibbs states. This is due to the coupling between diagonal and off-diagonal density matrix elements generated by the external driving. 

Let us denote the instantaneous Gibbs state corresponding to the Hamiltonian $H_\text{S}(t)$ with inverse temperature $\beta$ by {$\rho_\text{S}^\beta(t)=\exp(-\beta H_\text{S}(t))/Z_\text{S}(t)$} with partition function {$Z_\text{S}(t)=\Tr[\exp(-\beta H_\text{S}(t))]$}. As it is a Gaussian state, it is fully described by its first and second order moments given by $\expval{x}^\beta\!(t)=\lambda(t)$ and $\expval{p}^\beta\!(t)=0$, as well as $V_x^\beta(t)=V_p^\beta(t)=2n_\beta(\omega)+1$ and $C_{xp}^\beta(t)=0$. However, the instantaneous steady state $\rho^\text{ss}_{\text{S}}(t)$ of the nonadiabatic master equation \eqref{eq:nonad_me} satisfies the condition \cite{sca19}
\begin{equation}
	\mathcal{L}_t\left[\rho^\text{ss}_{\text{S}}(t)\right]=0.
	\label{eq:L_rho_th_zero}
\end{equation}
Instead of calculating $\rho_\text{S}^\text{ss}(t)$ directly, we use again that, like the instantaneous Gibbs state, it has to be a Gaussian state, and can thus be completely characterized by its first and second order moments. The latter can be obtained from the adjoint master equations \eqref{eq:adj_me_cm} by setting their left side to zero, solving for the complex moments, and afterwards computing the corresponding real moments. This leads to $\expval{x}^\text{ss}\!(t)=-2\Re[h(t)/\delta(\omega)]$ and $\expval{p}^\text{ss}\!(t)=-2\Im[h(t)/\delta(\omega)]$, as well as $V_x^\text{ss}(t)=V_p^\text{ss}(t)=2n_\beta(\omega)+1$ and $C_{xp}^\text{ss}(t)=0$. In the case of the adiabatic master equation, we again have to replace $h(t)$ by $h_\text{ad}(t)$. While the second order moments are the same as those for the instantaneous Gibbs state, this is clearly not the case for the first order moments, even for slow driving (since the driving term couples  diagonal and off-diagonal density matrix elements). As the instantaneous steady state is the state which the solution of the master equation should actually approach for slow driving, it becomes clear why the intuitive assumption of an approach to the instantaneous Gibbs state is wrong in general. 

To further gain deeper insight into the problem, we examine the fidelity between the actual state $\rho_\text{S}(t)$ and the instantaneous Gibbs state $\rho_\text{S}^\beta(t)$ on the one hand and  
between the actual state $\rho_\text{S}(t)$ and the instantaneous steady state $\rho_\text{S}^\text{ss}(t)$ on the other hand. The fidelity between two arbitrary single mode Gaussian states is given by \cite{ban15}
\begin{equation}
	\mathcal{F}(\rho_1,\rho_2)=\frac{1}{\sqrt{\Delta+\Lambda}-\sqrt{\Lambda}}\exp(-\frac{1}{2}\delta u^\mathrm{T}V^{-1}\delta u),
	\label{41}
\end{equation}
where we have defined $\delta u=u_2-u_1$ and $V=V_1+V_2$ with the displacement vector $u_i = (\expval{x}_i,\expval{p}_i)^\mathrm{T}$ and the covariance matrix of the $i$-th state
\begin{equation}
	V_i=
	\begin{pmatrix}
		V_x^i & C_{xp}^i\\
		C_{xp}^i & V_p^i
	\end{pmatrix}.
\end{equation} 
 Furthermore, we have introduced the quantities $\Delta=(1/4)\det(V)$ and $\Lambda=(1/4)\det(V_1+{i}M)\det(V_2+{i}M)$ with the matrix
\begin{equation}
	M=
	\begin{pmatrix}
		0 & 1\\
		-1 & 0
	\end{pmatrix}
	.
\end{equation}
Note that we have slightly modified the expression given in Ref.~\cite{ban15}: firstly, we adapted for different prefactors in the definition of dimensionless position and momentum operator and, secondly, we took the square of the quantity in Ref.~\cite{ban15}, which is
can still be interpreted as a fidelity and allows for better visibility of deviations of the fidelity from one.

The second row of \cref{fig:coherences} displays the fidelity between the actual state and the instantaneous Gibbs state (blue and orange lines) as well as the fidelity between the actual state and the instantaneous steady states of both master equations (light blue and red lines) for different driving times. For slow driving, the solutions of the master equations approach their instantaneous steady states, which are different from the corresponding instantaneous Gibbs states. As a result, the driven harmonic oscillator exhibits quantum coherence in the instantaneous energy eigenbases, but not in the eigenbases of the instantaneous steady states. Deviations between the system state and the instantaneous steady state, as well as between the predictions of the two master equations appear for fast driving. In particular, coherence is also seen in the eigenbases of the instantaneous steady states for both master equations, albeit much smaller than in the instantaneous energy eigenbasis. The latter coherence measure can be evaluated from Eq.~\eqref{eq:C_var_basis} by replacing $n_t$ by the number operator corresponding to the eigenbasis of the instantaneous steady 
state, which is computed in \cref{app_sec:n_t_in_inst_ss_basis}.

 \section{Conclusion}
\label{sec:conclusion}

We have derived a nonadiabatic master equation for a linearly driven harmonic oscillator, valid for generic, non-periodic driving. This Markovian master equation is of Lindblad form and thus ensures positivity.  We have explicitly solved the nonadiabatic master equation for a linear ramp and initial Gaussian states, and found that its predictions strongly depart from those of the corresponding adiabatic master equation, both for dynamical variables, like position and momentum, and thermodynamic quantities, like the energy -- the entropy is, however, the same in both instances for linear driving. We have additionally shown that the driven oscillator exhibits quantum coherence in the instantaneous energy eigenbasis for slow driving, and in both the  instantaneous energy eigenbasis and the eigenbasis of the instantaneous steady states for fast driving. This is related to the coupling between diagonal and nondiagonal density matrix elements generated by the external driving. In addition,  the state of the system, which approaches the instantaneous steady state in the limit of slow driving, differs from the instantaneous Gibbs state. The nonadiabatic master equation  allows one to describe fast quantum dynamics in regimes where adiabatic master equations are no longer applicable \cite{alb12,yip18}. It should therefore be useful to describe nonequilibrium quantum processes that occur arbitrarily far from equilibrium.

\section*{Acknowledgment}
The authors acknowledge support from the German Science Foundation DFG (Grant No. FOR 2724).

\newpage

\appendix

\begin{widetext}
	
\section{Interaction picture representation of creation and annihilation operators}
\label{app_sec:a_a_dagger_ip}

In this section, we derive the interaction picture representations of $a$ and $a^\dagger$. 
In general, it is a hard problem to find an exact expression for the unitary generated by a time-dependent Hamiltonian. However, it is possible, for example, if the Hamiltonian is a linear combination of the elements of closed Lie algebra, see Ref.~\cite{fer89} and references therein. Indeed, the operators $\mathds{1}$, $a$, $a^\dagger$, $n$, $a^2$ and $a^{\dagger 2}$ form a closed Lie-algebra, making it possible to derive a closed form expression for $U_\text{S}(t)$ \cite{kla93,aga70}. We here use a simpler method presented in Ref.~\cite{fer89}, where $\tilde{a}(t)$ and $\tilde{a}^\dagger(t)$ are calculated directly from the Heisenberg equation of motion. For a Hamiltonian that is no more than quadratic in $a$ and $a^\dagger$, i.e. a linear combination of elements of a closed Lie algebra, $\tilde{a}(t)$ and $\tilde{a}^\dagger(t)$ can be written as \cite{fer89}
\begin{subequations}
	\begin{align}
		\tilde{a}(t)&=v_0(t)+v_1(t)a+v_{2}(t)a^\dagger,\\
		\tilde{a}^\dagger(t)&=v_0^\ast(t)+v_2^\ast(t)a+v_1^\ast(t)a^\dagger.
	\end{align}
	\label{eq:a_a_dagger_ansatz}%
\end{subequations}%
The coefficient functions $u_i(t)$ can be determined by inserting Eq.~\eqref{eq:a_a_dagger_ansatz} into the Heisenberg equations of motion for $\tilde{a}(t)$ and $\tilde{a}^\dagger(t)$. This leads to the following system of differential equations 
\begin{subequations}
	\begin{align}
		\dot{v}_0(t)&={i}\omega v_0(t)-{i}\omega\frac{\lambda(t)}{2},\\
		\dot{v}_1(t)&=-{i}\omega v_1(t),\\		
		\dot{v}_2(t)&={i}\omega v_2(t),
	\end{align}
\end{subequations}%
with the initial conditions $v_0(0)=v_2(0)=0$ and $v_1(0)=1$. These differential equations can easily be solved yielding 
\begin{subequations}
	\begin{align}
	v_0(t)&={e}^{-{i}\omega t}A(t),\\
	v_1(t)&={e}^{-{i}\omega t},\\
	v_2(t)&=0, 
	\end{align}
\end{subequations}%
where $A(t)=\frac{{i}\omega}{2}\int_0^t{d}t^\prime\lambda(t^\prime)\exp({i}\omega t^\prime)$ in the main text. Inserting this into Eq.~\eqref{eq:a_a_dagger_ansatz} leads to Eq.~\eqref{eq:a_tilde}. This corresponds to the system time evolution operator 
\begin{equation}
	U_\text{S}(t)=\exp(-\mathrm{i}\omega t\,n)D(A(t)),
	\label{eq:explicit_form_of_U}
\end{equation}
where $D$ is the displacement operator. Alternatively one could have computed $U_\text{S}(t)$ directly using the Magnus series \cite{kla93} or phase space methods \cite{aga70}, which yields the same result up to irrelevant global phase factors.

\section{Damping coefficients, Lamb shift terms and driving function}
\label{app_sec:damp_coeff_driv_func}

In this section, we provide explicit expressions for the damping coefficients $\gamma_{12}(\omega)$ and $\gamma_{21}(\omega)$, for the Lamb shift terms $\Sigma_{12}(\omega)$ and $\Sigma_{12}(\omega)$ as well as for the driving function $f(t)$. For the calculation of all these quantities, we first need to compute the bath correlation functions $C_{ij}(s)=\langle \tilde{B}_i(s)B_j\rangle_\beta/\hbar^2$ with the bath operators $B_1=b=\sum_kg_kb_k$ and $B_2=b^\dagger$ and with $\expval{\dots}_\beta$ denoting the expectation value for the bath being in a thermal state $\rho_\text{B}^\beta$. Using $\tilde{b}_k(s)={e}^{-{i}\omega_ks}b_k$ and $\expval{b_kb_l}_\beta=\langle b_k^\dagger b_l^\dagger\rangle_\beta=0$, it can be easily seen that $C_{ii}(s)=0$. Furthermore, using $\langle b_k^\dagger b_l\rangle_\beta=\delta_{kl}\expval{n_k}_\beta$ and $\langle b_k b_l^\dagger\rangle_\beta=\delta_{kl}(\expval{n_k}_\beta+1)$ with $n_k=b_k^\dagger b_k$, we have  
\begin{subequations}
	\begin{align}
		C_{12}(s)&=\sum_k\left(\frac{|g_k|}{\hbar}\right)^2\mathrm{e}^{-{i}\omega_k s}(\expval{n_k}_\beta+1)=\int_0^\infty {d}\omega^\prime J(\omega^\prime)\left(n_\beta(\omega^\prime)+1\right){e}^{-{i}\omega^\prime s},\\
		C_{21}(s)&=\sum_k\left(\frac{|g_k|}{\hbar}\right)^2\mathrm{e}^{{i}\omega_k s}\expval{n_k}_\beta=\int_0^\infty {d}\omega^\prime J(\omega^\prime)n_\beta(\omega^\prime){e}^{{i}\omega^\prime s}.
	\end{align}
\end{subequations}%
In the second step in each line, in order to write the sums as integrals, we have introduced the spectral density $J(\omega)=\sum_{k}(|g_k|/\hbar)^2\delta(\omega-\omega_k)$ and used $\expval{n_k}_\beta=n_\beta(\omega_k)$ with $n_\beta(\omega)=\left[\exp(\beta\hbar\omega)-1\right]^{-1}$. As a next step, we write
\begin{subequations}
	\begin{align}
		\Gamma_{12}(\omega)&=\int_0^\infty{d}s\left(\int_0^\infty{d}\omega^\prime J(\omega^\prime)\left(n_\beta(\omega^\prime)+1\right){e}^{-{i}\omega^\prime s}\right){e}^{{i}\omega s}=\int_0^\infty{d}\omega^\prime J(\omega^\prime)\left(n_\beta(\omega^\prime)+1\right)\int_0^\infty{d}s{e}^{{i}(\omega-\omega^\prime)s}\nonumber\\
		&=\pi J(\omega)\left(n_\beta(\omega)+1\right)-{i}\mathcal{P}\int_0^\infty{d}\omega^\prime\frac{J(\omega^\prime)\left(n_\beta(\omega^\prime)+1\right)}{\omega^\prime-\omega}\equiv\frac{\gamma_{12}(\omega)}{2}+{i}\Sigma_{12}(\omega),\\
		\Gamma_{21}(\omega)&=\int_0^\infty {d}s\left(\int_0^\infty{d}\omega^\prime J(\omega^\prime)n_\beta(\omega^\prime){e}^{{i}\omega^\prime s}\right){e}^{-{i}\omega s}=\int_0^\infty{d}\omega^\prime J(\omega^\prime)n_\beta(\omega^\prime)\int_0^\infty{d}s{e}^{{i}(\omega^\prime-\omega)s}\nonumber\\
		&=\pi J(\omega)n_\beta(\omega)+{i}\mathcal{P}\int_0^\infty{d}\omega^\prime\frac{J(\omega^\prime)n_\beta(\omega^\prime)}{\omega^\prime-\omega}\equiv\frac{\gamma_{21}(\omega)}{2}+{i}\Sigma_{21}(\omega).
	\end{align}
\end{subequations}%
In the second last step of each line, we have used the identity $\int_0^\infty{d}s{e}^{{i}xs}=\pi\delta(x)+{i}\mathcal{P}(1/x)$, where $\mathcal{P}$ is the principal value \cite{kus06}. Furthermore, we introduce
\begin{subequations}
	\begin{align}
		\gamma(\omega)&=\gamma_{12}(\omega)-\gamma_{21}(\omega)=2\pi J(\omega),\\
		\Sigma(\omega)&=\Sigma_{12}(\omega)+\Sigma_{21}(\omega)=-\mathcal{P}\int_0^\infty{d}\omega^\prime\frac{J(\omega^\prime)}{\omega^\prime-\omega}.
	\end{align}
\end{subequations}%
The driving function $f(t)$ was originally defined as 
\begin{equation}
	f(t)=\int_0^t{d}s{\,C(s)}{e}^{{i}\omega s}A(t-s).
\end{equation}
We now calculate the difference of the two bath correlation functions using the spectral density introduced above and obtain
\begin{equation}
	{C(s)}=C_{12}(s)-C_{21}^\ast(s)=\sum_k\left(\frac{|g_k|}{\hbar}\right)^2{e}^{-{i}\omega_ks}=\int_0^\infty{d}\omega J(\omega){e}^{-{i}\omega s}.
	\label{eq:C_12_minus_C_21}
\end{equation}
Finally, we apply the continuum limit, where the discrete spectral density is approximated by a continuous function \cite{bre02,gar04,ali07,riv12}. A typical choice that we will also use here is the so called Ohmic spectral density with an exponential cutoff which reads $J(\omega)=\eta\omega\exp(-\omega/\Omega)$,
with the dimensionless coupling strength $\eta$ and the cutoff-frequency $\Omega$. With the Ohmic spectral density, we can solve the integral in Eq.~\eqref{eq:C_12_minus_C_21} analytically which finally enables us to write
\begin{equation}
	{C(s)=\frac{\eta}{\left(\frac{1}{\Omega}+{i}s\right)^2}.}
	\label{eq:f}
\end{equation}

\section{Master equation in the Schrödinger picture}
\label{app_sec:master_equation_Schroedinger}

In the following, we will explicitly show how to transform the interaction picture nonadiabatic master equation \eqref{eq:nonad_me_ip} from the main text into the Schrödinger picture. First, we note that the density matrix in the Schrödinger picture can be written as $\rho_\text{S}(t)=U_\text{S}(t)\tilde{\rho}_\text{S}(t)U_\text{S}^\dagger(t)$. This immediately yields
\begin{equation}
	\frac{d\rho_\text{S}(t)}{dt}=-\frac{{i}}{\hbar}\comm{H_\text{S}(t)}{\rho_\text{S}(t)}+U_\text{S}(t)\frac{d\tilde{\rho}_\text{S}(t)}{dt}U_\text{S}^\dagger(t).
\end{equation}
Inserting the master equation in the interaction picture, we then obtain
\begin{equation}
	U_\text{S}(t)\frac{d\tilde{\rho}_\text{S}(t)}{dt}U_\text{S}^\dagger(t)=-\frac{{i}}{\hbar}\comm{U_\text{S}(t)H_\text{LS}U_\text{S}^\dagger(t)}{\rho_\text{S}(t)}-\frac{{i}}{\hbar}\comm{U_\text{S}(t)\tilde{K}(t)U_\text{S}^\dagger(t)}{\rho_\text{S}(t)}+U_\text{S}(t)\mathcal{D}\left[\tilde{\rho}_\text{S}(t)\right]U_\text{S}^\dagger(t).
\end{equation}
Applying $U_\text{S}(t)\cdot U_\text{S}^\dagger(t)$ on Eq.~\eqref{eq:a_tilde}, multiplying it with $\exp({i}\omega t)$ and subtracting $A(t)$ we find
\begin{equation}
	U_\text{S}(t)aU_\text{S}^\dagger(t)={e}^{{i}\omega t}a-A(t).
\end{equation}
Using this, a lengthy but straightforward calculation yields
\begin{subequations}
	\begin{align}
		-\frac{{i}}{\hbar}\comm{U_\text{S}(t)H_\text{LS}U_\text{S}^\dagger(t)}{\rho_\text{S}(t) }&=-\frac{{i}}{\hbar}\comm{H_\text{LS}}{\rho_\text{S}(t)}-\frac{{i}}{\hbar}\comm{{i}\hbar\left({e}^{{i}\omega t}{i}\Sigma(\omega)A^\ast(t)a+{e}^{{-i}\omega t}{i}\Sigma(\omega)A(t)a^\dagger\right)}{\rho_\text{S}(t)},\\
		-\frac{{i}}{\hbar}\comm{U_\text{S}(t)\tilde{K}(t)U_\text{S}^\dagger(t)}{\rho_\text{S}(t) }&=-\frac{{i}}{\hbar}\comm{{i}\hbar\left({e}^{{i}\omega t}f^\ast(t)a-{e}^{-{i}\omega t}f(t)a^\dagger\right)}{\rho_\text{S}(t)},
	\end{align}
	\label{eq:H_LS_and_w_tilde_to_Schroedinger}%
\end{subequations}%
where we have used $\Sigma(\omega)=\Sigma_{12}(\omega)+\Sigma_{21}(\omega)$, as well as
\begin{equation}
	\begin{aligned}
		U_\text{S}(t)\mathcal{D}\left[\tilde{\rho}_\text{S}(t)\right]U_\text{S}^\dagger(t)&=\mathcal{D}\left[\rho_\text{S}(t)\right]-\frac{{i}}{\hbar}\comm{{i}\hbar\left(-{e}^{{i}\omega t}\frac{\gamma(\omega)}{2}A^\ast(t)a+{e}^{-{i}\omega t}\frac{\gamma(\omega)}{2}A(t)a^\dagger\right)}{\rho_\text{S}(t)},
	\end{aligned}
	\label{eq:diss_to_Schroedinger}
\end{equation}
with $\gamma(\omega)=\gamma_{12}(\omega)+\gamma_{21}(\omega)$. Putting everything together leads to
\begin{equation}
	U_\text{S}(t)\frac{d\tilde{\rho}_\text{S}(t)}{dt}U_\text{S}^\dagger(t)=-\frac{{i}}{\hbar}\comm{H_\text{LS}}{\rho_\text{S}(t)}+\mathcal{D}\left[\rho_\text{S}(t)\right]-\frac{{i}}{\hbar}\comm{{i}\hbar\left(g^\ast(t)a-g(t)a^\dagger\right)}{\rho_\text{S}(t)},
\end{equation}
where $g(t)=\exp({i}\omega t)[f(t)-\alpha(\omega)A(t)]$ with $\alpha(\omega)=\gamma(\omega)/2+{i}\Sigma(\omega)$, which immediately leads to the Schrödinger picture master equation \eqref{eq:nonad_me} from the main text.

\section{Recovering the weakly driven master equation}
\label{app_sec:wd_me}

In this section, we analyze the weakly driven master equation in more detail. This equation is often applied in quantum optics and in the study of dissipative quantum systems \cite{bre02,gar04,ali07,riv12,car93,wei08}, and provides a simple description in which the driving enters the master equation only in the Hamiltonian term. It can be derived from the nonadiabatic master equation \eqref{eq:nonad_me} by writing the system Hamiltonian as $H_\text{S}(t)=H_{\text{S}}(0)+\varepsilon H_{\text{D}}(t)$, where $H_{\text{S}}(0)$ is a constant part and $H_{\text{D}}(t)$ represents the driving Hamiltonian with amplitude $\varepsilon$. In the limit of small $\varepsilon$, the time evolution operator may be expanded as 
\begin{equation}
	U_\text{S}^\text{wd}(t)= {e}^{ {i}tH_{\text{S}}(0)/\hbar}+\mathcal{O}(\varepsilon).
	\label{eq:U_weak_driving}
\end{equation}
Keeping only the lowest order in the driving amplitude and in the weak bath coupling {regime}, one obtains \cite{liu18}
\begin{eqnarray}
	\frac{d{\rho}_\text{S}(t)}{dt}&=&-\frac{ {i}}{\hbar}\comm{H_\text{S}(t)+H_\text{LS}}{\rho_\text{S}(t)}+\mathcal{D}\left[\rho_\text{S}(t)\right] \nonumber \\&=&{\mathcal{L}}_{t,\text{wd}}\left[{\rho}_\text{S}(t)\right],
	\label{eq:wd_me}
\end{eqnarray}
in the Schrödinger picture representation. The adjoint master equations take the same form as for the nonadiabatic master equation, i.e. Eqs.~\eqref{eq:ad_me}, with the only difference that $h(t)$ is replaced by 
$h_\text{wd}(t)=-\frac{\mathrm{i}\omega}{2}\lambda(t)$
for the weakly driven case.

As mentioned in the main text, second order moments of the solution of the master equation are independent of the driving, and therefore are the same for all three master equations including the weakly driven master equation \eqref{eq:wd_me}. As a consequence, it suffices to investigate the behavior of the first order moments of the solution. In \cref{fig:x_and_p_wd_1} we compare the predictions made by the nonadiabatic master equation \eqref{eq:nonad_me} and the weakly driven master equation \eqref{eq:wd_me} for the difference of the expectation value of the dimensionless position to the minimum o
the harmonic potential $\expval{x}(\tau)-\lambda(\tau)$ (upper row), as well as the expectation value of the dimensionless momentum $\expval{p}(\tau)$ (lower row), as functions of the dimensionless time $\tau$. The initial state, the driving speeds and all other parameters were chosen as in \cref{fig:x_and_p}.

\begin{figure}[t]
	\includegraphics[width=\textwidth]{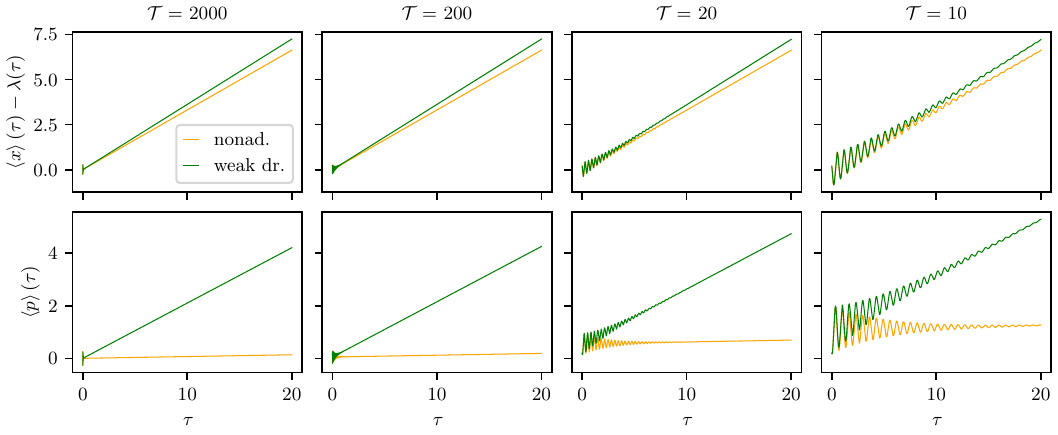}
	\caption{The figure displays the difference of the expectation value of the dimensionless position to the minimum of the harmonic potential $\expval{x}(\tau)-\lambda(\tau)$ (upper row) as well as the expectation value of the dimensionless momentum $\expval{p}(\tau)$ (lower row) as functions of the dimensionless time $\tau$ predicted by the nonadiabatic master equation \eqref{eq:nonad_me} (green line) and the weakly driven master equation \eqref{eq:ad_me} (orange line) for different driving speeds: very slow ($\mathcal{T}=2000$, first column), slow ($\mathcal{T}=200$, second column), intermediate ($\mathcal{T}=20$, third column) and quick driving ($\mathcal{T}=10$, fourth column). The other parameters are fixed at the values $y=0.1$, $w=4$, $\eta=0.008$, $\Delta l=10$ and the initial state is determined by $\expval{a}(0)=0.1+0.1{i}$, $V(0)=0$ and $C(0)=n_\text{th}+\delta n_0-|\expval{a}(0)|^{\raisebox{-2pt}{\scriptsize 2}}$ with $\delta n_0=2$.}
	\label{fig:x_and_p_wd_1}
\end{figure}

Note that the qualitative behavior predicted by the weakly driven master equation is roughly the same as for the other two master equations: {for position and momentum} one can observe a linear increase superposed by a decaying oscillation, the amplitude of which increases with the driving speed. While, in contrast to the adiabatic master equation, the weakly driven master equation slightly overestimates the slope of the linear increase in position, we still have an overall acceptable agreement with the nonadiabatic master equation. For the momentum, however, the slope of the linear increase is massively overestimated by the weakly driven master equation. 

We want to stress  that, at least for sufficiently long transient time, the deviation between both master equations seemingly does not notably depend on the driving speed. Only for very short transient times, it seems as if the weakly driven master equation becomes even better for increasing driving speeds. However, this is mainly an artifact of the scaling of the transient time: we parametrize the driving speed by the total duration of the protocol ($T$ or $\mathcal{T}$), while keeping initial and endpoint of the protocol or, equivalently, its strength ($\Delta l$) fixed. Thus, quick driving automatically means short driving here. As our transient time $\tau=t/T$ is scaled with the duration, a fixed interval in $\tau$ corresponds to a long interval in $t$ for slow driving and to a short interval in $t$ for quick driving. Thus, if the $\tau$-interval with good agreement between both master equations increases with the driving speed, this does not mean that the $t$-interval with good agreements does so too.
Overall, we can conclude that the deviation of the weakly driven master equation is small for small transient times and increases rapidly with increasing transient time. However, we observed only a weak dependence on the driving speed. The previous observation is not surprising, {as} in the derivation of the weakly driven master equation, contrary to the adiabatic master equation, no explicit assumption on the driving speed was made. Instead, as its name already suggests, a weak driving was presumed. It is hence worthwhile to examine how the results presented in \cref{fig:x_and_p_wd_1} change if we decrease the driving strength $\Delta l$.

\begin{figure}[t]
	\includegraphics[width=\textwidth]{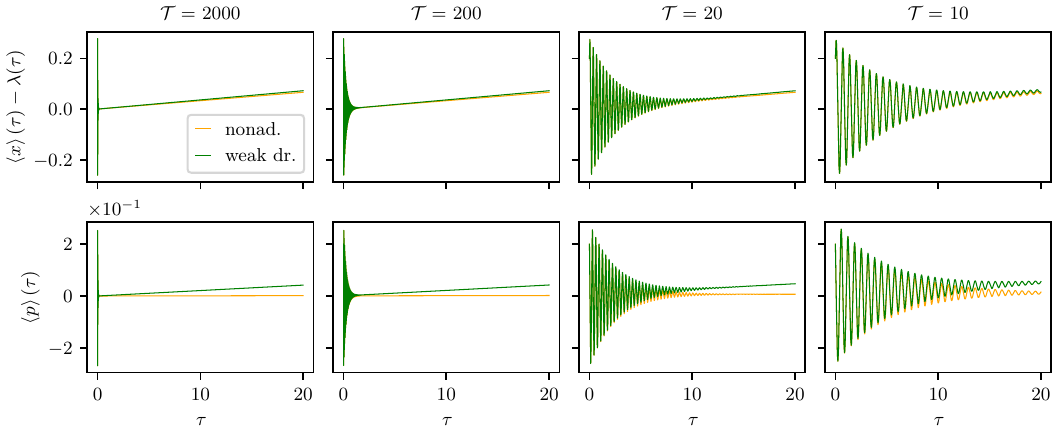}
	\caption{The figure displays the difference of the expectation value of the dimensionless position to the minimum of the harmonic potential $\expval{x}(\tau)-\lambda(\tau)$ (upper row) as well as the expectation value of the dimensionless momentum $\expval{p}(\tau)$ (lower row) as functions of the dimensionless time $\tau$ predicted by the nonadiabatic master equation \eqref{eq:nonad_me} (green line) and the weakly driven master equation \eqref{eq:ad_me} (orange line) for different driving speeds: very slow ($\mathcal{T}=2000$, first column), slow ($\mathcal{T}=200$, second column), intermediate ($\mathcal{T}=20$, third column) and quick driving ($\mathcal{T}=10$, fourth column). The other parameters are fixed at the values $y=0.1$, $w=4$, $\eta=0.008$, $\Delta l=0.1$ and the initial state is determined by $\expval{a}(0)=0.1+0.1{i}$, $V(0)=0$ and $C(0)=n_\text{th}+\delta n_0-|\expval{a}(0)|^{\raisebox{-2pt}{\scriptsize 2}}$ with $\delta n_0=2$.}
	\label{fig:x_and_p_wd_2}
\end{figure}
  
In \cref{fig:x_and_p_wd_2} we change the driving strength to a weak driving by setting $\Delta  l=0.1$. The qualitative behavior is in principle the same as for strong driving, however, the slope of the linear increase of position and momentum is significantly lower, as expected. Although at least the results for the momentum predicted by the two master equations slowly diverge over time, the maximal deviation is small, and we have a good overall agreement, showing that the weakly driven master equation indeed can be considered a valid approximation for sufficiently weak driving.

\section{Details of the adiabatic master equation}
\label{app_sec:details_ad_me}

In this section, we provide some further details for the derivation of the adiabatic master equation for the harmonic oscillator. First we denote that system Hamiltonian can be written as 
\begin{equation}
	H_\text{S}(t)=\hbar\omega D\left(\frac{\lambda(t)}{2}\right)H_\text{S}(0)D^\dagger\left(\frac{\lambda(t)}{2}\right),
	\label{eq:H_S_new}
\end{equation}
where $D$ is the displacement operator. Since $H_\text{S}(0)=\hbar\omega(n+1/2)$, its instantaneous energy eigenvalues and eigenstates are respectively given by $\varepsilon_n(t)=\hbar\omega(n+1/2)=\varepsilon_n$ and $\ket{\varepsilon_n(t)}=D(\lambda(t)/2)\ket{n}$.  Using the derivative of the displacement operator, we can show that the derivative of the eigenstate is given by 
\begin{equation}
	\ket{\dot{\varepsilon}_n(t)}=\frac{\dot{\lambda}(t)}{2}\left(a^\dagger-a\right)\ket{\varepsilon_n(t)},
\end{equation}
which yields a vanishing Berry connection $\phi_n(t)= {i}\hbar \braket{\varepsilon_n(t)}{\dot{\varepsilon}_n(t)}=0$ and therefore $\mu_n(t)=\varepsilon_n t$. We can then explicitly write the adiabatic time evolution operator as 
\begin{equation}
	U_\text{S}^\text{ad}(t)=\exp(-\frac{\mathrm{i}\omega t}{2})\sum_n\exp(-\mathrm{i}\omega tn)\dyad{\varepsilon_n(t)}{n}.
\end{equation}
If we apply the latter on the annihilation operator $a$, we obtain
\begin{equation}
	\tilde{a}_\text{ad}(t)={e}^{{i}\omega t}(A_\text{ad}(t)+a),
	\label{eq:a_tilde_ad}
\end{equation}
with $A_\text{ad}(t)=[\lambda(t)/2]\exp({i}\omega t)$. Using again the system Hamiltonian in the form of Eq.~\eqref{eq:H_S_new}, we find 
\begin{align}
	{e}^{{i}sH_\text{S}(t)/\hbar}=D\left(\frac{\lambda(t)}{2}\right){e}^{{i}sH_\text{S}(0)\hbar}D^\dagger\left(\frac{\lambda(t)}{2}\right).
\end{align}
From $\tilde{O}_\text{S}^\text{ad}(t-s)=\bar{U}_\text{S}^{\text{ad}\dagger}(t,s)O\bar{U}_\text{S}^\text{ad}(t,s)$ with  $\bar{U}_\text{S}^\text{ad}(t,s)=\exp({ {i}sH_\text{S}(t)/\hbar})U_\text{S}^\text{ad}(t)$ it then follows that
\begin{equation}
	\tilde{a}_\text{ad}(t-s)={e}^{-{i}\omega (t-s)}(A_\text{ad}(t,s)+a), 
	\label{eq:a_tilde_ad_2}
\end{equation}
where $A_\text{ad}(t,s)=[\lambda(t)/2]\exp({i}\omega(t-s))$. As soon as one has obtained the expressions \eqref{eq:a_tilde_ad} and \eqref{eq:a_tilde_ad_2}, one can derive the adiabatic master equation in complete analogy with the nonadiabatic one, starting from Eq.~\eqref{eq:driven_Redfield_HO} with the only difference that $\tilde{a}(t)$ and $\tilde{a}(t-s)$ are replaced by $\tilde{a}_\text{ad}(t)$ and $\tilde{a}_\text{ad}(t-s)$, respectively.

\section{Integrals in the driving function}
\label{app_sec:integrals}

In this section, we provide full analytic expressions as well as the asymptotic behavior of the integrals \eqref{eq:integrals} appearing in the driving function. For the first integral, we have
\begin{equation}
	I_0(z)=\int_0^z{d}x\frac{1}{\left(1+{i}x\right)^2}=\int_1^{1+{i}z}{d}y\frac{-{i}}{y^2}=\left[\frac{i}{y}\right]_1^{1+iz}=\frac{{i}}{1+{i}z}-{i}=\frac{1}{z-{i}}-{i}.
\end{equation}
The second integral is given by
\begin{equation}
	\begin{aligned}
		I_1(z)&=\int_0^z{d}x\frac{x}{\left(1+{i}x\right)^2}=\int_1^{1+{i}z}{d}y\frac{1-y}{y^2}=\left[-\frac{1}{y}\right]_1^{1+{i}z}-\left[\ln(y)\right]_1^{1+{i}z}=1-\frac{1}{1+{i}z}-\ln(1+{i}z)\\
		&=\frac{z}{z-{i}}-\ln(z-{i})-{i}\frac{\pi}{2},
	\end{aligned}
\end{equation}
where we have used the main branch of the complex logarithm. Finally, for the third integral, we find 
\begin{equation}
	\begin{aligned}
		I_\text{e}(z)&=\int_0^z{d}x\frac{{e}^{{i}x/w}}{\left(1+{i}x\right)^2}=\int_1^{1+{i}z}{d}y\frac{-{i}{e}^{(y-1)/w}}{y^2}=-{i}{e}^{-1/w}\int_1^{1+iz}{d}y\frac{{e}^{y/w}}{y^2}=-{i}(1/w){e}^{-1/w}\int_{1/w}^{1/w+iz/w}{d}u\frac{{e}^u}{u^2}\\
		&\equiv A\int_a^b{d}u\frac{e^u}{u^2}=A\left(\left[-\frac{e^u}{u}\right]_a^b+\int_a^b{d}u\frac{e^u}{u}\right)=A\left(\frac{e^a}{a}-\frac{e^b}{b}+\mathrm{Ei}(b)-\mathrm{Ei}(a)\right)\\
		&=-{i}(1/w){e}^{-1/w}\left(\frac{{e}^{1/w}}{1/w}-\frac{{e}^{1/w+{i}z/w}}{1/w+{i}z/w}+\mathrm{Ei}(1/w+{i}z/w)-\mathrm{Ei}(1/w)\right)\\
		&=({i}/w){e}^{-1/w}\left(\mathrm{Ei}(1/w)-\mathrm{Ei}(1/w+{i}z/w)\right)+\frac{{e}^{{i}z/w}}{z-{i}}-{i},
	\end{aligned}
\end{equation}
where $\mathrm{Ei}(x)$ is the exponential integral \cite{abr72}. Now that we found analytic expression for all three integrals, we analyze their asymptotic behavior. For this purpose, we first use that the terms $\frac{1}{z-{i}}$ and $\frac{{e}^{{i}z/w}}{z-{i}}$ quickly converge to zero and that the term $\frac{z}{z-{i}}$ quickly converges to one for increasing $z$. Furthermore, we write $\ln(z-{i})=\ln(|z-{i}|)+{i}\arg(z-{i})=\frac{1}{2}\ln(1+z^2)+{i}\arctan(-1/z)$, with the imaginary part quickly converging to zero for increasing $z$ while the real part diverges. Finally, using the limit $\lim_{b\to\infty}\mathrm{Ei}(a+{i}b)={i}\pi$ \cite{abr72} one can see that the asymptotic behavior of the three integrals is given by
\begin{subequations}
	\begin{align}
		\bar{I}_0&=-{i},\\
		\bar{I}_1(z)&=1-\frac{1}{2}\ln(1+z^2)-{i}\frac{\pi}{2},\\
		\bar{I}_\text{e}&=({i}/w){e}^{-1/w}\left(\mathrm{Ei}(1/w)-{i}\pi\right)-{i}.
	\end{align}
\end{subequations}%
The divergence of $\bar{I}_1(z)$ for $z\to\infty$ is the reason we cannot set the upper integration limit in $f(t)$ to infinity. Finally, it is worthwhile to connect the asymptotic values of the integrals with the damping coefficients and the Lamb shift terms of the undriven amplitude damping master equation. One finds
\begin{equation}
	\begin{aligned}
		\alpha(\omega)&=\frac{\gamma(\omega)}{2}+{i}\Sigma(\omega)=\frac{\gamma_{12}(\omega)-\gamma_{21}(\omega)}{2}+{i}\left(\Sigma_{12}(\omega)-\Sigma_{21}(\omega)\right)=\Gamma_{12}(\omega)-\Gamma_{21}^\ast(\omega)\\
		&=\int_0^\infty{d}s\left(C_{12}(s)-C_{21}^\ast(s)\right){e}^{{i}\omega s}=\eta\int_0^\infty{d}s\frac{{e}^{{i}\omega s}}{\left(\frac{1}{\Omega}+{i}s\right)^2}=\eta\Omega^2\int_0^\infty{d}s\frac{{e}^{{i}\omega s}}{\left(1+{i}\Omega s\right)^2}\\
		&=\omega\eta w\int_0^\infty{d}x\frac{{e}^{{i}x/w}}{\left(1+{i}x\right)^2}=\omega\eta w\bar{I}_\text{e}.
	\end{aligned}
\end{equation}
In complete analogy, we can show that $\alpha(0)=\omega\eta w\bar{I}_0$.

\section{Solving the master equation explicitly}
\label{app_sec:solve_me}

In the main text, we did not solve any of the master equations \eqref{eq:nonad_me} and \eqref{eq:ad_me} explicitly, as it is {significantly} easier to calculate desired observables directly via the adjoint master equations. For completeness, we provide here the recipe to solve the master equations themselves as well. We concretely present two possible approaches to this problem. The first approach, being a slight generalization to the one presented in \cite{jie03} aims to solve the master equation in full generality by reducing the solution to the one of the undriven master equation for which a full solution is known but complicated \cite{arn96,hon10}. The second approach employs an ansatz used in Ref.~\cite{jie03} to solve the undriven master equation and has the appeal of being quite simple although not fully general.

For the first approach, it is convenient to note that both master equations \eqref{eq:nonad_me} and \eqref{eq:ad_me} can be written in the form
\begin{equation}
	\dot{\rho}=-\frac{{i}}{h}\comm{H_\text{S}+H_\text{LS}+K}{\rho}+\mathcal{D}\left[\rho\right]=\comm{-{i}(\omega+\Sigma)n+h^\ast a-ha^\dagger}{\rho}+\mathcal{D}\left[\rho\right],
\end{equation}
where we omitted all arguments and subscripts that are not strictly necessary. Now, we introduce the density matrix
\begin{equation}
	\rho^\prime=D(\zeta)\rho D^\dagger(\zeta).
\end{equation}
Using the Kermack-McCrae form of the displacement operator \cite{ger23},
\begin{equation}
	D(\zeta)=\exp(\frac{1}{2}|\zeta|^2)\exp(-\zeta^\ast a)\exp(\zeta a^\dagger),
\end{equation}
using the product rule, we find that its derivative is given by 
\begin{equation}
	\dot{\left(D(\zeta)\right)}=\left(\frac{\dot{\zeta}^\ast\zeta}{2}-\frac{\zeta^\ast\dot{\zeta}}{2}-\dot{\zeta}^\ast a+\dot{\zeta}a^\dagger\right)D(\zeta).
	\label{eq:dot_D}
\end{equation}
This immediately yields
\begin{equation}
	\dot{\rho}^\prime=\comm{-\dot{\zeta}^\ast a+\dot{\zeta}a^\dagger}{\rho^\prime}+D(\zeta)\dot{\rho}D^\dagger(\zeta).
\end{equation}
We can then show that
\begin{equation}
	D(\zeta)\comm{-{i}(\omega+\Sigma)n+h^\ast a-ha^\dagger}{\rho}D^\dagger(\zeta)=\comm{-{i}(\omega+\Sigma)n+(h-{i}(\omega+\Sigma)\zeta)^\ast a-(h-{i}(\omega+\Sigma)\zeta) a^\dagger}{\rho^\prime}.
\end{equation}
Furthermore, we have
\begin{equation}
	\begin{aligned}
		D(\zeta)\mathcal{D}\left[\rho\right]D^\dagger(\zeta)&=\gamma_{12}\left((a-\zeta)\rho^\prime(a^\dagger-\zeta^\ast)-\frac{1}{2}\acomm{(a^\dagger-\zeta^\ast)(a-\zeta)}{\rho^\prime}\right)\\
		&+\gamma_{21}\left((a^\dagger-\zeta^\ast)\rho^\prime(a-\zeta)-\frac{1}{2}\acomm{(a-\zeta)(a^\dagger-\zeta^\ast)}{\rho^\prime}\right)\\
		&=\mathcal{D}\left[\rho^\prime\right]+\comm{\left(-\frac{\gamma}{2}\zeta\right)^\ast a-\left(-\frac{\gamma}{2}\zeta\right)a^\dagger}{\rho^\prime}.
	\end{aligned}
\end{equation}
Summarizing everything, we obtain
\begin{equation}
	\dot{\rho}=-\frac{{i}}{\hbar}\comm{H_\text{S}+H_\text{LS}}{\rho^\prime}+\mathcal{D}\left[\rho^\prime\right]+\comm{X^\ast a-Xa^\dagger}{\rho^\prime},
\end{equation}
where we have defined the quantity 
\begin{equation}
	X=-\dot{\zeta}-\left(\frac{\gamma}{2}+{i}(\Sigma+\omega)\right)\zeta+h.
\end{equation}
Thus, if $\chi$ is chosen such that $X=0$, the master equation for $\rho^\prime$ is exactly the one without driving. There exists extensive literature on how to solve the undriven master equation in full generality \cite{arn96,hon10}. However, the corresponding solution is usually complicated, which is why we will not discuss this any further in this work. Instead, we next present  a second approach to the problem which significantly simpler, albeit not as general.
	
The second approach is a generalization of the method presented in Ref.~\cite{muf93} using the ansatz
\begin{equation}
	\rho={e}^\phi{e}^{\alpha a^\dagger}{e}^{\chi n}{e}^{\alpha^\ast a},
	\label{eq:Mufti_ansatz}
\end{equation}
where $\phi$ and $\chi$ are real functions, while $\alpha$ is a complex function. The ansatz \eqref{eq:Mufti_ansatz} is a possible parametrization of a single mode Gaussian state. Although it is not the most general single mode Gaussian state, its appeal lies in its simplicity, and it is the most general solution for all initial states that can be cast in the same form. In Ref.~\cite{muf93} the ansatz \eqref{eq:Mufti_ansatz} was used to solve the undriven master equation of a harmonic oscillator. Here, we generalize it to the driven case. First, we note that the inverse of Eq.~\eqref{eq:Mufti_ansatz} can be written as
\begin{equation}
	\rho^{-1}={e}^{-\phi}{e}^{-\alpha^\ast a}{e}^{-\chi n}{e}^{-\alpha a^\dagger}.
	\label{eq:Mufti_ansatz_inv}
\end{equation}
Using the relation ${e}^{A}B{e}^{-A}=\sum_{k=0}^\infty\frac{1}{k!}\comm{A}{B}_k$ with $\comm{A}{B}_k=\comm{A}{\comm{A}{B}_{k-1}}$ and $\comm{A}{B}_0=B$, we can show that 
\begin{subequations}
	\begin{align}
		\rho a\rho^{-1}&={e}^{-\chi}(a-\alpha),\\
		\rho a^\dagger\rho^{-1}&={e}^\chi a^\dagger+\alpha^\ast,\\
		\rho n\rho^{-1}&=n-\alpha a^\dagger+\alpha^\ast{e}^{-\chi}a-|\alpha|^2{e}^{-\chi}.
	\end{align}
	\label{eq:rho_something_rho_inv}%
\end{subequations}%
We recall that all master equations we wish to solve can be cast in the form
\begin{equation}
	\dot{\rho}\rho^{-1}=-\frac{{i}}{h}\comm{H_\text{S}+H_\text{LS}+w}{\rho}\rho^{-1}+\mathcal{D}\left[\rho\right]\rho^{-1}.
	\label{eq:generic_me_times_rho_inv}
\end{equation}
With the help of Eq.~\eqref{eq:rho_something_rho_inv} we find that the left side of Eq.~\eqref{eq:generic_me_times_rho_inv} is given by
\begin{equation}
	\dot{\rho}\rho^{-1}=\left(\dot{\phi}-{e}^{-\chi}\alpha\dot{\alpha}^\ast\right)+\mathrm{e}^{-\chi}\dot{\alpha}^\ast a+\left(\dot{\alpha}-\alpha\dot{\chi}\right)a^\dagger+\dot{\chi}n.
	\label{eq:rho_dot}
\end{equation}
Furthermore, for the first term on the right side of Eq.~\eqref{eq:generic_me_times_rho_inv} we find
\begin{equation}
	\begin{aligned}
		&-\frac{{i}}{h}\comm{H_\text{S}+H_\text{LS}+w}{\rho}\rho^{-1}=\comm{-{i}(\omega+\Sigma)n+h^\ast a-ha^\dagger}{\rho}\rho^{-1}\\
		&=-{i}(\omega+\Sigma)\left(n-\rho  n\rho^{-1}\right)+h^\ast\left(a-\rho a\rho^{-1}\right)-h\left(a^\dagger-\rho a^\dagger\rho^{-1}\right)\\
		&=\left(-{i}(\omega+\Sigma)|\alpha|^2{e}^{-\chi}+h^\ast{e}^{-\chi}\alpha+h\alpha^\ast\right)+\left({i}(\omega+\Sigma)\alpha^\ast{e}^{-\chi}+h^\ast(1-{e}^{-\chi})\right)a+\left(-{i}(\omega+\Sigma)\alpha+h({e}^\chi-1)\right)a^\dagger
	\end{aligned}
\end{equation}
and second term of Eq.~\eqref{eq:generic_me_times_rho_inv} reads
\begin{equation}
	\begin{aligned}
		&\mathcal{D}\left[\rho\right]\rho^{-1}=\gamma_{12}\left(a\rho a^\dagger-\frac{1}{2}\acomm{a^\dagger a}{\rho}\right)\rho^{-1}+\gamma_{21}\left(a^\dagger\rho a-\frac{1}{2}\acomm{aa^\dagger}{\rho}\right)\rho^{-1}\\
		&=\gamma_{12}a\rho a^\dagger\rho^{-1}+\gamma_{21}a^\dagger\rho a\rho^{-1}-\sigma n-\sigma \rho n\rho^{-1}-\gamma_{21}\\
		&=\left(-\gamma_{21}+\gamma_{12}{e}^\chi+\sigma|\alpha|^2{e}^{-\chi}\right)+\alpha^\ast\left(\gamma_{12}-\sigma{e}^{-\chi}\right)a+\alpha\left(\sigma-\gamma_{21}{e}^{-\chi}\right)a^\dagger+\left(\gamma_{12}{e}^\chi+\gamma_{21}{e}^{-\chi}-2\sigma\right)n,
	\end{aligned}
	\label{eq:diss_regrouped}	
\end{equation}
where $\sigma=(\gamma_{12}+\gamma_{21})/2$. Now, we insert everything into Eq.~\eqref{eq:generic_me_times_rho_inv} and use the linear independence of the operators $\mathds{1}$, $a$, $a^\dagger$ and $n$ to equate the prefactors in front of them on each side of the equation This leads to the following system of differential equations for $\phi$, $\alpha$ and $\chi$:
\begin{subequations}
	\begin{align}
		\dot{\phi}-{e}^{-\chi}\alpha\dot{\alpha}^\ast&=-\gamma_{21}+\gamma_{12}{e}^\chi+\left(\sigma-{i}(\Sigma+\omega)\right)|\alpha|^2{e}^{-\chi}+h\alpha^\ast+h^\ast\alpha{e}^{-\chi},\\
		{e}^{-\chi}\dot{\alpha}^\ast&=\alpha^\ast\left(\gamma_{12}-\left(\sigma-{i}(\Sigma+\omega)\right){e}^{-\chi}\right)+h^\ast\left(1-{e}^{-\chi}\right),\\
		\dot{\alpha}-\alpha\dot{\chi}&=\alpha\left(\left(\sigma-{i}(\Sigma+\omega)\right)-\gamma_{21}{e}^{-\chi}\right)+h\left({e}^\chi-1\right),\\
		\dot{\chi}&=\gamma_{12}\mathrm{e}^\chi+\gamma_{21}{e}^{-\chi}-2\sigma.
	\end{align}
	\label{eq:diff_eqs_Mufti}%
\end{subequations}%
Inserting the fourth into the third equation yields
\begin{equation}
	\dot{\alpha}=\left(\gamma_{12}{e}^\chi-\left(\sigma+{i}(\omega+\Sigma)\right)\right)\alpha+h\left({e}^\chi-1\right).
\end{equation}	
This is the complex conjugate of the second equation multiplied by ${e}^\chi$. Thus, the second equation can be discarded. By inserting the second equation into the first equation, the latter can be simplified to  
\begin{equation}
	\dot{\phi}=\gamma_{12}\left(|\alpha|^2+{e}^\chi\right)-\gamma_{21}+h^\ast\alpha+h\alpha^\ast.
\end{equation}
We do not need to solve this equation, though, as the $\phi$ can be obtained by normalization. In Ref.~\cite{muf93} it was shown that the condition $\Tr(\rho)=1$ yields 
\begin{equation}
	{e}^\phi=\left(1-z\right)\exp(\frac{|\alpha|^2}{z-1}),
	\label{eq:phasefactor}
\end{equation}
where we have introduced the new variable $z={e}^\chi$ with $\dot{z}=\dot{\chi}z$. With the latter expression, Eq.~\eqref{eq:diff_eqs_Mufti} finally reduces to
\begin{subequations}
	\begin{align}
		\dot{z}&=-2\sigma z+\gamma_{12}z^2+\gamma_{21},\\
		\dot{\alpha}&=\left(\gamma_{12}z-\left(\sigma+{i}(\Sigma+\omega)\right)\right)\alpha+h(z-1).
	\end{align}
	\label{eq:diff_eqs_Mufti_simplifies}%
\end{subequations}%
A recipe on how observables can be calculated from this ansatz can be extracted from Ref.~\cite{muf93}.

\section{Dimensionless quantities needed for the numerics}
\label{app_sec:dimless}

In this section, we make the dimensionless quantities $\bar{\delta}$ and $\bar{\gamma}$ as well as $\bar{h}(\tau)$ and $\bar{h}_\text{ad}(\tau)$ introduced in Eq.~\eqref{eq:a_and_V_and_C_dimensionless} in main text explicit. First, we have
\begin{subequations}
	\begin{align}
		\bar{\delta}&=T\delta(\omega)=T\alpha(\omega)+{i}\omega T\equiv \bar{\alpha}+{i}\mathcal{T},\\
		\bar{\alpha}&=T\alpha(\omega)=\frac{T\gamma(\omega)}{2}+{i}T\Sigma(\omega)\equiv\frac{\bar{\gamma}}{2}+{i}\bar{\Sigma}.
	\end{align}
\end{subequations}%
Furthermore, we have defined
\begin{subequations}
	\begin{align}
		\bar{\gamma}&=T\gamma(\omega)=2\pi TJ(\omega)=2\pi\eta\omega T{e}^{-\frac{\omega}{\Omega}}=2\pi\eta\mathcal{T}{e}^{-\frac{1}{w}},\\
		\bar{\Sigma}&=T\Sigma(\omega)=-\mathcal{P}\int_0^\infty{d}\omega^\prime\frac{TJ(\omega^\prime)}{\omega^\prime-\omega}=-\mathcal{P}\int_0^\infty{d}x\frac{TJ(\omega x)}{x-1}=-\mathcal{P}\int_0^\infty{d}x\frac{\eta\mathcal{T} x{e}^{-\frac{x}{w}}}{x-1}.
	\end{align}
\end{subequations}%
Next, we find for the driving functions
\begin{subequations}
	\begin{align}
		\bar{h}(\tau)&=Th(T\tau)=Tg(T\tau)-\frac{{i}\omega T}{2}\lambda(T\tau)\equiv \bar{g}(\tau)-\frac{{i}\mathcal{T}}{2}\bar{\lambda}(\tau),\\
		\bar{h}_\text{ad}(\tau)&=Th_\text{ad}(T\tau)=Tg_\text{ad}(T\tau)-\frac{{i}\omega T}{2}\lambda(T\tau)-\frac{T}{2}\dot{\lambda}(T\tau)\equiv \bar{g}_\text{ad}(\tau)-\frac{{i}\mathcal{T}}{2}\bar{\lambda}(\tau)-\frac{1}{2}\dot{\bar{\lambda}}(\tau),
	\end{align}
\end{subequations}%
with $\bar{\lambda}(\tau)=\lambda(T\tau)=\Delta l\tau$ and 
\begin{subequations}
	\begin{align}
		\bar{g}(\tau)&=Tg(T\tau)={e}^{-{i}\mathcal{T}\tau}\left(Tf(T\tau)-T\alpha(\omega)A(T\tau)\right)={e}^{-{i}\mathcal{T}\tau}\left(\bar{f}(\tau)-a\bar{A}(\tau)\right),\\
		\bar{g}_\text{ad}(\tau)&=Tg_\text{ad}(T\tau)={e}^{-{i}\mathcal{T}\tau}\left(Tf_\text{ad}(T\tau)-T\alpha(\omega)A_\text{ad}(T\tau)\right)={e}^{-{i}\mathcal{T}\tau}\left(\bar{f}_{ad}(\tau)-a\bar{A}_\text{ad}(\tau)\right).
	\end{align}
\end{subequations}%
Hereby, we have introduced 
\begin{subequations}
	\begin{align}
		\bar{f}(\tau)&=Tf(\tau)=(\eta\Delta l/2)\left(\left[\left(w\mathcal{T}\tau+ {i}w\right)I_0(w\mathcal{T}\tau)-I_1(w\mathcal{T}\tau)\right]\exp( {i}\mathcal{T}\tau)-{i}wI_\text{e}(w\mathcal{T}\tau)\right),\\
		\bar{f}_\text{ad}(\tau)&=Tf_\text{ad}(\tau)=(\eta\Delta l/2)\left(w\mathcal{T}\tau\exp( {i}\mathcal{T}\tau)I_0(w\mathcal{T}\tau)\right)
	\end{align}
\end{subequations}%
with the integrals $I_0(z)$, $I_1(z)$ and $I_\text{e}(z)$ defined as in Eqs.~\eqref{eq:integrals} and with
\begin{subequations}
	\begin{align}
		\bar{A}(\tau)&=A(T\tau)=\frac{{i}\omega}{2}\int_0^{T\tau}{d}t^\prime\lambda(t^\prime){e}^{{i}\omega t^\prime}=\frac{{i}\mathcal{T}}{2}\int_0^{\tau}{d}\tau^\prime\bar{\lambda}(\tau^\prime){e}^{{i}\mathcal{T}\tau^\prime}=\frac{\Delta l}{2}\left(\left(\tau+\frac{{i}}{\mathcal{T}}\right){e}^{{i}\mathcal{T}\tau}-\frac{{i}}{\mathcal{T}}\right),\\
		\bar{A}_\text{ad}(\tau)&=A_\text{ad}(\tau)=\frac{\lambda(T\tau)}{2}{e}^{{i}\mathcal{T}\tau}=\frac{\Delta l}{2}\tau{e}^{{i}\mathcal{T}\tau}.
	\end{align}
\end{subequations}%
Finally, with $n_\beta(\omega)=[\exp(\beta\hbar\omega)-1]^{-1}=[\exp(1/y)-1]^{-1}= n_\text{th}$, we managed to express everything as functions of the dimensionless parameters $y$, $w$, $\eta$, $\mathcal{T}$ and $\Delta l$.

\section{Number operator in the eigenbasis of the instantaneous steady state}
\label{app_sec:n_t_in_inst_ss_basis}

In this section, we derive the number operator of the eigenbasis of the instantaneous steady state $\rho_\text{S}^\text{ss}(t)$. First, we note that the instantaneous steady state is a single mode Gaussian state. Any such state can be written as a displaced squeezed thermal state
\begin{equation}
	\rho=D(\alpha)S(\xi)\frac{{e}^{-\nu n}}{Z}S^\dagger(\xi)D^\dagger(\alpha),
	\label{eq:DSTS}
\end{equation}
where we have defined the displacement operator $D(\alpha)=\exp(\alpha a^\dagger-\alpha^\ast a)$ and the squeezing operator $S(\xi)=\exp(\frac{1}{2}\xi^\ast a^2-\frac{1}{2}\xi a^{\dagger 2})$ with $\xi=r{e}^{{i}\theta}$, as well as the partition function $Z=\Tr({e}^{-\nu n})=\left[1-{e}^{-\nu}\right]^{-1}$ as usual.  One can then show that the complex first and second order moments of this state are given by
\begin{subequations}
	\begin{align}
		\expval{a}&=\alpha,\\
		V_a&=PQ(2N(\nu)+1),\\
		C_{aa^\smalldagger}&=R^2+\left(P^2+R^2\right)N(\nu),
	\end{align}
	\label{eq:moments_DSTS}%
\end{subequations}%
with $P=\cosh(r)$, $Q=-{e}^{{i}\theta}\sinh(r)$, $R=|Q|=\sinh(r)$ and $N(\nu)=\left[{e}^\nu-1\right]^{-1}$. Rearranging these equations, we obtain
\begin{subequations}
	\begin{align}
		R^2&=(2C_{aa^\smalldagger}+1- \mu)/2\mu,\\
		P^2+R^2&=(2C_{aa^\smalldagger}+1)/\mu,\\
		PQ&=V_a/\mu,
	\end{align}
	\label{eq:moments_DSTS_inverted}%
\end{subequations}%
with $\mu=\sqrt{(2C+1)^2-4|V|^2}$. Next, we use that the state in Eq.~\eqref{eq:DSTS} can be rewritten as 
\begin{equation}
	\rho=\frac{{e}^{-\nu \tilde{n}}}{\Tr({e}^{-\nu \tilde{n}})},
\end{equation}
where we have introduced the displaced squeezed number operator
\begin{equation}
	\tilde{n}=D(\alpha)S(\xi)nS^\dagger(\xi)D^\dagger(\alpha).
\end{equation}
This is the number operator corresponding to the basis $\tilde{\mathcal{B}}=\lbrace\ket{\tilde{n}}\rbrace_{n=0}^\infty$ with $\ket{\tilde{n}}=D(\alpha)S(\xi)\ket{n}$, i.e. $\tilde{n}\ket{\tilde{n}}=n\ket{\tilde{n}}$. Obviously, $\rho$ is diagonal in this basis. What remains to be done is to find a more explicit expression for $\tilde{n}$. Using the properties of displacement and squeezing operators, we find
\begin{equation}
	\begin{aligned}
		\tilde{n}&=\left(P^2+R^2\right)n-PQ a^{\dagger 2}-PQ^\ast a^2+\left(2PQ\alpha^\ast-\left(P^2+R^2\right)\alpha\right)a^\dagger+\left(2PQ^\ast\alpha-\left(P^2+R^2\right)\alpha^\ast\right)a\\
		&+\left(P^2+R^2\right)|\alpha|^2-PQ\alpha^{\ast 2}-PQ^\ast\alpha^2+R^2.
	\end{aligned}
\end{equation}
Exploiting Eq.~\eqref{eq:moments_DSTS_inverted} then yields
\begin{equation}
	\begin{aligned}
		\mu(\tilde{n}+1/2)-(C_{aa^\smalldagger}+1/2)&=(2C_{aa^\smalldagger}+1)n-V_aa^{\dagger 2}-V_a^\ast a^2+(2V_a\alpha^\ast-(2C_{aa^\smalldagger}+1)\alpha)a^\dagger+(2V_a^\ast\alpha-(2C_{aa^\smalldagger}+1)\alpha^\ast)a\\
		&+(2C_{aa^\smalldagger}+1)|\alpha|^2-V_a\alpha^{\ast 2}-V_a^\ast\alpha^2.
	\end{aligned}
\end{equation}
Finally, we want to express everything by the real first and second order moments of Eq.~\eqref{eq:DSTS} and by Hermitian operators. Using $\expval{x}=2\Re(\expval{a})$ and $\expval{p}=2\Im(\expval{a})$ as well as $V_x=2C_{aa^\smalldagger}+1+2\Re(V_a)$, $V_p=2C_{aa^\smalldagger}+1-2\Re(V_a)$  and $C_{xp}=2\Im(V_a)$, we arrive at the expression
\begin{equation}
	\begin{aligned}
		\mu(\tilde{n}+1/2)&=\frac{V_p}{4}x^2+\frac{V_x}{4}p^2+\frac{C_{xp}}{4}\acomm{x}{p}+\frac{1}{2}\left(C_{xp}\expval{p}-V_p\expval{x}\right)x-\frac{1}{2}\left(C_{xp}\expval{x}-V_x\expval{p}\right)p\\
		&+\frac{V_p}{4}\expval{x}^2+\frac{V_x}{4}\expval{p}^2+\frac{C_{xp}}{2}\expval{x}\!\!\expval{p}.
	\end{aligned}
\end{equation}
We have now expressed the number operator $\tilde{n}$ corresponding to the basis in which the most general single mode Gaussian state $\rho$ is diagonal as a linear combination of $x$, $p$, $x^2$, $p^2$ and $\acomm{x}{p}$, where the coefficients contain only the real first and second order moments of $\rho$. Thus, to obtain the number operator corresponding to the basis in which the instantaneous steady state $\rho_\text{S}^\text{ss}(t)$ is diagonal, we just have to use the real first and second order moments of $\rho_\text{S}^\text{ss}(t)$ instead.

\end{widetext}


\begin{thebibliography}{99}
\bibitem{HSrxp} This can be easily seen by writing the system Hamiltonian as $H_\text{S}(t)=(p_0p)^2/2m+m\omega^2x_0^2[x-\lambda(t)]^2/2$, where $x_0=\sqrt{\hbar/2m\omega}$ and $p_0=\sqrt{\hbar m\omega/2}$ are the he zero-point fluctuations of the harmonic oscillator in position and momentum and
$x=a^\smalldagger + a$ and $p={i}(a^\smalldagger-a)$ are the dimensionless position and momentum operator.
\end{thebibliography}
\end{document}